\documentclass[a4paper,10pt]{article}

 \usepackage[cp1251]{inputenc}

\usepackage{cite,amsmath,amsfonts,amsthm,fullpage}
\usepackage{youngtab}

\newcommand{\sgn}{\mathop\mathrm{sgn}\nolimits}
\newcommand{\Pa}{\mathop\mathrm{P}\nolimits}

\newcommand{\bpow}{\mathbf{p}}
\newcommand{\bbpow}{{\bar{\mathbf{p}}}}

\theoremstyle{plain}

\newtheorem{Lemma}{Lemma}
\newtheorem{Proposition}{Proposition}
\newtheorem{Corollary}{Corollary}
\newtheorem{Remark}{Remark}

\theoremstyle{remark}

\def\l{\langle}
\def\r{\rangle}

\def\Tr{\mathrm {Tr}}
\def\tr{\mathrm {tr}}
\def\det{\mathrm {det}}

\def\res{\mathop{\mathrm {res}}\limits_}

\def\bp{\begin{Proposition}}
\def\ep{\end{Proposition}}
\def\bc{\begin{Corollary}}
\def\ec{\end{Corollary}}
\def\bl{\begin{Lemma}}
\def\el{\end{Lemma}}
\def\be{\begin{equation}}
\def\ee{\end{equation}}
\def\br{\begin{Remark}\rm\small}
\def\er{\end{Remark}}
\def\brs{\begin{remarks}.\\ \rm\
\begin{enumerate}}
\def\ers{\end{enumerate}\end{remarks}}
\def\bea{\begin{eqnarray}}
\def\eea{\end{eqnarray}}

\def\Tr{\mathrm {Tr}}
\def\tr{\mathrm {tr}}
\def\det{\mathrm {det}}

\def\sgn{\mathrm {sgn}}

\def\res{\mathop{\mathrm {res}}\limits}

\def\&{&{\hskip -20pt}}

\newcount\YDcount\YDcount=0
\def\YDsize{10pt}

\def\YD#1{%
\ifnum#1=0
 \ifnum\YDcount=0 \ifx\varnothing\undefined\emptyset\else\varnothing\fi
 \else\vskip1.4pt\egroup\YDcount=0\fi
\else
 \ifnum\YDcount=0 \YDcount=1\vcenter\bgroup\vskip1pt
 \else\nointerlineskip\fi
 \vbox{\hrule\hbox{\vrule height\YDsize
 \loop\hskip\YDsize\vrule\ifnum\YDcount<#1\advance\YDcount1\repeat}\hrule
 \kern-0.4pt}\expandafter\YD
\fi}

\usepackage[usenames,dvipsnames]{color}
\usepackage{ulem}

\begin{document}

\author{ Sergei M. Natanzon\thanks{National Research University Higher School of Economics, Moscow, Russia; Institute for
Theoretical and Experimental Physics, Moscow, Russia;
email: natanzons@mail.ru} \and Aleksandr Yu.
Orlov\thanks{Institute of Oceanology, Nahimovskii Prospekt 36,
Moscow 117997, Russia, and National Research University Higher School of Economics,
International Laboratory of Representation
Theory and Mathematical Physics,
20 Myasnitskaya Ulitsa, Moscow 101000, Russia, email: orlovs@ocean.ru
}}
\title{BKP and projective Hurwitz numbers}

\maketitle

\begin{abstract}

We consider $d$-fold branched coverings of the projective plane $\mathbb{RP}^2$ and show that the hypergeometric tau function
 of the BKP hierarchy of Kac and van de Leur is the generating function for weighted sums of the related Hurwitz numbers.
In particular we get the $\mathbb{RP}^2$ analogues of the $\mathbb{CP}^1$ generating functions proposed
by Okounkov and by Goulden and Jackson. Other examples are Hurwitz numbers weighted by the Hall-Littlewood and by the Macdonald polynomials.
We also consider integrals of tau functions which generate projective Hurwitz numbers and Hurwitz numbers related to
different Euler characterisitics of the base Klein surfaces.

\end{abstract}

\bigskip


\textbf{2010 Mathematic Subject Classification:} 05A15, 14N10, 17B80, 35Q51, 35Q53, 35Q55, 37K20, 37K30,

\section{Introduction}

In  \cite{Okounkov-2000} A. Okounkov studied
ramified coverings of the Riemann sphere, having arbitrary ramification
type over $0$ and $\infty$, and with simple ramifications elsewhere, and  made the
seminal observation that the
generating function for the related Hurwitz numbers (numbers of nonequivalent coverings with
given ramification type) is a tau function for the Toda lattice hierarchy. Later  links
between the study of covers and integrable system were further developed by Okounkov and Pandharipande
\cite{Okounkov-Pand-2006} and by Goulden and Jackson \cite{Goulden-Jackson-2008}.
Then a number of papers concerning the topic was written \cite{MM1},
\cite{ AMMN-2011, AMMN-2014},\cite{DuninKazarian---2013},\cite{Zog},
\cite{Harnad-2014},\cite{HO-2014},\cite{Harnad-October-2014}.
The review of this topic may be found in \cite{Uspehi-KazarianLando} and in \cite{Harnad-overview-2015}.
For the most recent works see \cite{ACEH-2016-11} and references therein.

On the other hand it was shown that certain matrix models also generate
Hurwitz numbers \cite{Zog}, \cite{GGPN}, \cite{KZ}, \cite{HarnadMathieu-sept-2014},\cite{Chekhov-2014}.
It is not so surprising since tau functions used for generating of Hurwitz numbers belong to a special family 
 found in \cite{KMMM} and in \cite{OS-2000},\cite{OS-TMP} which are called tau functions of
\textit{hypergeometric type},
and such tau functions were used as asymptotic expansion of matrix integrals in \cite{HO-Borel},\cite{O-2002},
\cite{OShiota-2004},\cite{HO-2MM}. Hypergeometric tau functions are multivariable generalizations of hypergeometric
series where the role of Gauss hypergeometric equation plays the so-called string equation \cite{OS-2000} and
matrix integrals may be viewed as an analogue of the integral representation of Gauss hypergeometric series.

All works on Hurwitz numbers cited above are devoted to the counting of covers of the Riemann sphere and
the links of this problem to the Toda lattice (TL) and Kadomtsev-Petviashvili (KP) hierarchies.

The covering problem of the Riemann sphere (Euler characteristic $\textsc{e}=2$) goes back to classical results by Frobenius and
Schur \cite{F,FS}. We can refer the readers to the wonderful
textbook \cite{ZL} where the general case of the enumeration of covers of Riemann surfaces of higher genus was considered.
The Frobenius-type formula for the Hurwitz numbers enumerating $d$-fold branched coverings of connected Riemann or Klein
surfaces (without boundary) of any Euler characteristic $\textsc{e}$
was obtained in the papers of A. Mednykh and G. Pozdnyakova \cite{M1},\cite{M2}  and also Gareth A. Jones \cite{GARETH.A.JONES}.
 It contains the sum over irreducible representations
$\lambda$ of the symmetric group $S_d$ (see  \cite{F,FS,M1,M2,ZL,GARETH.A.JONES})
\be\label{Hurwitz-counting}
H^{\textsc{e},\textsc{f}}(d; \Delta^{(1)}\,\dots,\Delta^{(\textsc{f})})\,=\,
\sum_{\lambda} \,\,\left(\frac{{\rm dim}\lambda}{d!}\right)^\textsc{e}
\,
\prod_{i=1}^\textsc{f} \,\varphi_\lambda(\Delta^{(i)})\,,
\ee
where $\textsc{e}$ is the Euler characteristic of the base surface $\Omega$,
$\Delta^{(i)}$ are profiles  over branch points on $\Omega$,
${\rm dim}\lambda$ is the dimension of the irreducible representation of $S_d$, and
\be
\label{varphi}
\varphi_\lambda(\Delta^{(i)}) := |C_{\Delta^{(i)}}|\,\,\frac{\chi_\lambda(\Delta^{(i)})}{{\rm dim}\lambda} ,
\quad {\rm dim}\lambda:=\chi_\lambda\left((1^d)\right)
\ee
$\chi_\lambda(\Delta)$ is the character of the symmetric group $S_d$ evaluated at a cycle type $\Delta$,
and $\chi_\lambda$ ranges over the irreducible complex characters of $S_d$,
labeled by partitions $\lambda=(\lambda_1,\dots,\lambda_{\ell})$. The convenient notion of normalized character,
$\varphi_\lambda$, comes from \cite{Okounkov-2000},\cite{AMMN-2011}. Each
profile $\Delta^{(i)}$ is a partition of $d$ - the set of non-negative non-increasing numbers
$(d^{(i)}_1,d^{(i)}_2,\dots )$, which describes the ramification over the point number $i$ on the base. The weights
of all partitions involved in (\ref{Hurwitz-counting}) are equal:
$|\lambda|:=\sum_j\lambda_j=|\Delta^{(i)}|:=\sum_j d^{(i)}_j=d$. The number $|C_{\Delta}|$
is the number of elements in the cycle class $\Delta$ in $S_d$.
The sum (\ref{Hurwitz-counting}) runs over partitions of the weight $d$. We assume that
$\varphi_\lambda(\Delta)$ vanishes in case $|\lambda|\neq |\Delta|$.

The Hurwitz numbers form a topological field theory \cite{Dijkgraaf}. In the string theory applications the covering
surface is the worldsheet of the string while the base surface is the target space.
Hurwitz numbers are used in mathematical physics
(for instance in \cite{Dijkgraaf}) and in algebraic geometry \cite{ZL}. A lot of interest and the developments in
these studies appeared due to the work \cite{ELSV} which relates Hurwitz numbers to Gromov-Witten theory.

Our paper deals with enumeration of the covers of the projective plane $\mathbb{RP}^2$ (the case $\textsc{e}=1$ in
(\ref{Hurwitz-counting}));
the related Hurwitz numbers will be called projective. The projective Hurwitz numbers were introduced
in Mednykh-Pozdnyakova work \cite{M2} and independently in the context
of topological field theory in \cite{AN}.

In this
case we found that
it is a different hierarchy of integrable equations which is related to the problem: this is the BKP hierarchy
introduced by V.Kac and J. van de Leur in \cite{KvdLbispec}\footnote{This BKP hierarchy was called ``charged'' and ``fermionic''
BKP hierarchy in \cite{KvdLbispec}. We call it ``large'' BKP hierarchy because it includes KP one and may be related
\cite{LeurO-2014} to the two-component KP. The ``small'' KP
hierarchy, introduced in \cite{JM} is a subhierarchy in the KP one.}. In certain sense this hierarchy is very similar to the
DKP one introduced in \cite{JM}, however the difference between D and B types is crucial for the counting problem
we need (see Remark \ref{DKPvsBKP} in the Appendix).
Somehow the BKP hierarchy of Kac-van de Leur is not well-known, though it has applications to the famous
orthogonal and symplectic ensembles \cite{L1} and some other models of random matrices
and random partitions \cite{OST-I, O-2012, LeurO-2014}.
We are going to show that the BKP tau function of the hypergeometric type introduced in \cite{OST-I} generates Hurwitz
numbers for covers of $\mathbb{RP}^2$. Up to an unimportant factor the BKP tau function of the hypergeometric type may be
written as follows
\be\label{BKP-hyp-tau}
\tau^{\rm B}(N,n,\bpow)\,=\,\sum_{\lambda\in\Pa\atop\ell(\lambda)\le N}\,s_\lambda(\bpow)\,
c^{|\lambda|}\prod_{(i,j)\in\lambda}r(n+j-i)
\ee
where $s_\lambda$ is the Schur function \cite{Mac}, related to a partition $\lambda=(\lambda_1,\dots,\lambda_{\ell})$,
$\ell(\lambda)$ denotes the number of nonvanishing parts of $\lambda$, $c$ is a parameter and
 $\Pa$ denotes the set of all partitions (in what follows we will omit to point out the summation range $\Pa$,
 having it in mind).
The product in the right hand side ranges over all nodes of the Young diagram $\lambda$, $j$ is the column and $i$ is the row
coordinate of the node of $\lambda$ depicted  in English way where the diagonal spreads down and right from the origin.
Two discrete parameters $N$ and $n$ and the set $\bpow=(p_1,p_2,\dots)$
are called the BKP higher times \cite{KvdLbispec}\footnote{ In the present paper we use the so-called power sums
$p_m$ \cite{Mac} as higher time variables rather than $\frac 1m p_m$ as it is common in the soliton theory \cite{JM}.}.
We suppose that the tau function (\ref{BKP-hyp-tau}) is equal to 1 if $N=0$ and vanishes if $N<0$.
$r$ is an arbitrary chosen function of one variable, it will be specified later according to the needs of our work.
The number $j-i$ is called the \textit{content} of the node located at $i$-th row and $j$-th column of the Young diagram
$\lambda$; the product over all nodes of the Young diagram in the right hand side of (\ref{BKP-hyp-tau}) is called
 \textit{content product} (the generalized Pochhammer symbol). Content products play an essential role in the study of
 applications of the symmetric groups (for instance, see \cite{Goulden-Jackson-2008},\cite{HarnadMathieu-sept-2014},
\cite{Harnad-2014} and references therein). The special role of the content product in the study of Hurwitz numbers
generated by KP hierarchy was observed and worked out in \cite{GJ}.

In the present paper we chose two different types of parameterizations of the function $r$ which defines the content product
in (\ref{BKP-hyp-tau}). The first is
\be\label{r-MirMorNat}
\qquad({\rm I})\qquad\qquad\quad
r(x)= \,\exp \sum_{m>0}\,\frac 1m \zeta_m h^m x^m\,
\qquad\qquad\qquad\qquad
\ee
The second is
\be\label{r-new-trig}
\,({\rm II})\qquad\qquad\quad
r(x)= \, \texttt{t}^{x \xi_0 }\exp \sum_{m\neq 1}\,\frac 1m \xi_m \texttt{t}^{mx}\,
\qquad\qquad\qquad
\ee
The complex number $\texttt{t}$ and sets $\{\zeta_m,\,m > 0\}$ and $\{\xi_m,\,m\in\mathbb{Z} \}$
are free parameters. Similar to \cite{Okounkov-2000},\cite{Harnad-2014} we introduce auxiliary 
parameters $c$ and $h$, the powers of $c$ count the degree of covering maps while the powers of the
parameter $\frac 1h$ which enters (\ref{r-MirMorNat}) count the Euler characteristic
of the covers.
In what follows we may put $c=1$ and $h=1$ in cases we are not interested in the degree and the Euler characteristic,
and hope this does not produce a confusion. Everywhere if it is possible without a confusion we also avoid writing down
the dependence of $r(x)$ and of $\tau(N,n,\bpow)$ on $c,h,\zeta, \xi$ and other parameters to make formulae more readable.

Let us note that the usage of the parametrization I in the applications the content product was
also considered in \cite{Harnad-October-2014} in the study the combinatorial Hurwitz numbers by using of
Cayley graphs and Jusys-Murphy elements as it was suggested by Canadien combinatorial school \cite{GGPN} and
developed in \cite{Harnad-2014}.

One of the results of our paper is the explicit expressions of the content products parameterized by (\ref{r-MirMorNat}) and
by (\ref{r-new-trig}) in terms of characters of symmetric group, see Propositions \ref{Proposition-texttt-H}
and \ref{Proposition-texttt-T}.

 Let us write down the answer for the case (II):
\be\label{content-Introduction}
\prod_{(i,j)\in\lambda}r(x+j-i)= \texttt{t}^{\xi_0 x |\lambda|+\xi_0 \varphi_\lambda(\Gamma)}\exp
\sum_{m\neq 0}\, \frac1m \xi_m \texttt{t}^{mx} D_{p_1}\log s_\lambda(\bpow)|_{\bpow(0,\texttt{t}^m)},\quad |\Gamma|=|\lambda|=d
\ee
Here first we apply the Euler operator $D_{p_1}=p_1\frac{\partial}{\partial p_1}$  to the Schur function $s_\lambda(\bpow)$
where $\bpow=(p_1,p_2,\dots)$, then
evaluate the result at the point $\bpow=\bpow(0,\texttt{t}^m)=(p_1(0,\texttt{t}^m),p_2(0,\texttt{t}^m),\dots)$
where $p_k(0,\texttt{t}^m)=(1-\texttt{t}^{mk})^{-1}$.
The partition $\Gamma$ is defined as follows. For $d\ge 2 $ it is the partition $(1^{d-2}2)$,
 it's length is $\ell(\Gamma)=d-1$.  We choose the notation $\Gamma$ because the Young diagram of the
partition $(1^{d-2}2)$ resembles the capital Greek letter gamma. The cycle class labeled by $\Gamma$ in $S_d$
consists of all transpositions. We keep the notation $\Gamma$ also for the case $d\le 1$, then
$\Gamma=(d)$.

One can see that the content product for a partition $\lambda$ is expressed in terms of the Schur functions labeled
by the same partition. Thanks to the characteristic map relation \cite{Mac}
\be\label{Schur-char-map}
s_\lambda(\bpow)\,=\,\frac{{\rm dim}\lambda}{d!}
\left(p_1^d +\sum_{\Delta\atop \Delta\neq 1^d}\,\varphi_\lambda(\Delta) \bpow_\Delta  \right)
\ee
formula $(\ref{content-Introduction})$ produces series in $\varphi_\lambda$ which, in turn, due to the summation
over partitions $\lambda$ in (\ref{BKP-hyp-tau}), allows us to consider (\ref{BKP-hyp-tau}) as the generating function
for Hurwitz numbers (\ref{Hurwitz-counting}). The content product (\ref{content-Introduction}) is expressed
in terms of the Schur function and formula (\ref{Schur-char-map}) exhibits the explicit dependence of the Schur
functions on ${\rm dim \lambda}$. However in the expression for the content product (\ref{content-Introduction}) the
dependence on ${\rm dim \lambda}$ disappears thanks to the logarithmic derivative of the Schur function.
Then it follows that in the generating
series (\ref{Hurwitz-counting}) the exponent $\textsc{e}$ (the Euler characteristic) is equal to 1. Therefore we obtain
generating series for the projective Hurwitz numbers (to be precise - generating series for weighted sums of the projective Hurwitz
numbers, where as we shall see later, the weights are defined by specifications of the parameters $\xi,\texttt{t}$).

The case (I) may be considered as the degeneration of the case (II) when $\texttt{t}\to 1$, therefore
in this case series (\ref{BKP-hyp-tau}) also generate projective Hurwitz numbers.

Here and below, the notation $\bpow_\Delta$ serves for the product $p_{d_1}p_{d_2}\cdots$ where $d_i$ are the parts
of the partition $\Delta$: $\Delta=(d_1,d_2,\dots)$.

Then the tau function (\ref{BKP-hyp-tau}) may be presented as
\be\label{textsc-H}
\tau^{\rm B}(N,n,\bpow)=\sum_{d\ge 0} c^d \sum_{\Delta\atop |\Delta|=d}  \textsc{H}_r(d;\Delta) \bpow_\Delta,
\qquad\textsc{H}_r(d;\Delta)=\sum_{\lambda\atop |\lambda|=d,\,\ell(\lambda)\le N} \frac{{\rm dim}\lambda}{d!} \varphi_\lambda(\Delta)
\prod_{(i,j)\in\lambda} r(n+j-i)
\ee
where (for $d\le N$) $H_r(d;\Delta)$ is a certain series in Hurwitz numbers which describe $d$-fold covers with the ramification $\Delta$ over a
point, say, $0$ of $\mathbb{RP}^2$ and
ramifications over additional points which are determined by the choice of $r$, namely, of the parameters in (\ref{r-MirMorNat})
or in (\ref{r-new-trig}). We should keep in mind that in (\ref{BKP-hyp-tau}) it is only the part of sum over $\lambda$ 
conditioned by $|\lambda|\le N$ generates Hurwitz numbers $H^{\textsc{e},\textsc{f}}(d=|\lambda|;\dots)$. Thus to get Hurwitz 
number for the study of $d$-fold coverings one should work with series (\ref{BKP-hyp-tau}) conditioned by $N \ge d$. The same 
restriction we meet in Section \ref{Matrix-integrals} when consider integrals over $N\times N$ matrices which generate Hurwitz numbers.

\br
As we see, the sum
\be\label{E=0}
  \sum_{\lambda}\,c^{|\lambda|}\prod_{(i,j)\in\lambda}r(n+j-i)
\ee
also may be viewed as the generating function of the Hurwitz numbers when the base surface has Euler characteristic
equal to zero (which either the torus, or the Klein bottle). In case of the specification (\ref{r-MirMorNat}) such sums may be
related to the characters of the
Lie algebra of differential operators on the circle
as it was studied in \cite{BlochOkounkov}.
\er

Both choices of the content products, (\ref{r-MirMorNat}) and (\ref{r-new-trig}), contain the direct analogue of the
Okounkov generating series \cite{Okounkov-2000},
now, however for covers of $\mathbb{RP}^2$. It is enough to put $\zeta_m=0$ except $\zeta_1$ in (\ref{r-MirMorNat}),
or, to put all $\xi_m=0$ except $\xi_0$ in (\ref{r-new-trig}).

Using respectively (\ref{r-MirMorNat}) and (\ref{r-new-trig})  we obtain two different types of generating functions
for the projective Hurwitz numbers. The first one arising from (\ref{r-MirMorNat})
may be compared to the approach based on completed cycles developed in  \cite{Okounkov-Pand-2006},\cite{AMMN-2011}
(where the $\mathbb{CP}^1$
case was studied).
The second one, obtained from (\ref{r-new-trig}),  is related to a '$q$-deformation'
of the previous case (where instead of $q$ we use the letter $\texttt{t}$) which in turn may be compared to the
approaches
developed independently in \cite{NO-2014} and in \cite{Harnad-October-2014}. We will show that in the "$\texttt{t}$-deformed"
(or, "trigonometric") case the Hall-Littlewood and the Macdonald polynomials naturally appear as weight functions in
weighted sums of the projective Hurwitz numbers.

The structure of the paper is as follows. In Section \ref{definitions-section} we explain the notion of Hurwitz numbers for
Klein surfaces. In the Subsection \ref{Remarks-on-Mednykh-subsection} we present links between Hurwitz numbers with different Euler characteristics
$\textsc{e}$ of base surfaces. There we also explain the special role of the ramification described by the one row Young diagram
$(d)$ (\textit{maximally ramified profile}) in the enumeration
of the $d$-fold covers presented in Proposition \ref{d-cycle-proposition}. It means that BKP hypergeometric function
also generate Hurwitz numbers for the $d$-fold covers of any Klein surfaces in case at least one of the profiles is
maximally ramified.
In Section \ref{Content products} we find the content products for cases
(I) and (II) in terms of characters of the symmetric groups. The answers are respectively given by Propositions \ref{Proposition-texttt-H}
and \ref{Proposition-texttt-T}.\footnote{The Propositions \ref{Proposition-texttt-H} is actually a new version of the known
results
presented in \cite{AMMN-2011} about completed cycles, however we did not yet write down the correspondence in an explicit way.}
In Section \ref{Hurwitz-numbers-section} we introduce weighted sums of
the projective Hurwitz numbers (which as we will show in further sections are generated by the BKP tau functions). For the weighting
in particular
we use the Macdonald, Jack and Hall-Littlewood polynomials which
naturally appear via specifications of the parametrization of $r$. Now we are ready to use tau functions.
In Section \ref{BKP-tau-function} we recall the notion of the BKP hierarchy and of the special family of the BKP tau functions
called hypergeometric ones. We show that the BKP hypergeometric tau function may be obtained from the two-component KP hypergeometric
tau function (which may be related to the semiinfinite TL equations) by an action of a special heat operator which was introduced in
the Subsection \ref{Remarks-on-Mednykh-subsection}. This action
relates hierarchy serving respectively $\mathbb{CP}^1$ and $\mathbb{RP}^2$ Hurwitz counting problems.
At the end of this section we how we get hypergeometric BKP tau functions with content products (\ref{r-MirMorNat}) and
(\ref{r-new-trig}) in terms of an
action of vertex operators on a special tau function $\tau^{\rm B}_1$.
Section \ref{Examples-tau-subsection} is devoted to the set of examples of hypergeometric tau functions which are related to
different choices of the parameters in (\ref{r-MirMorNat}) and (\ref{r-new-trig}).
In Section \ref{BKP tau function as the generating function} our main results are written down. We show that the
tau function (\ref{BKP-hyp-tau}) together with either (\ref{r-MirMorNat}) or (\ref{r-new-trig}) generates projective Hurwitz numbers and
weighted sums of the projective Hurwitz numbers. We show that, choosing the content product as in
(\ref{r-new-trig}), sums weighted by Hall-Littlewood polynomials with the parameter $\texttt{t}$ appear naturally.
We present the BKP tau functions which generate Hurwitz numbers with an arbitrary profile over $0$ and additional
branch points with two types of profiles: maximally ramified profiles and minimally ramified profiles (the simple branch points).

In the last Section \ref{Matrix-integrals} we present certain integrals over matrices, which generate projective Hurwitz
numbers. Let us note that the well-known $\beta=2$ ensemble (or, the same, the unitary ensemble, or, the same, one-matrix model)
counts both $\mathbb{CP}^1$ Hurwitz numbers and ribbon graphs with a given number of faces, vertices and edges, see
\cite{MelloKochRamgoolam},
and as it was shown in \cite{NO-2014} (see Section 6 there) the simplest way to get it is to present the one-matrix model as a
hypergeometric tau function \cite{HO-2003}. We do not manage to do the same in $\mathbb{RP}^2$ case. The analogues of the
unitary ensemble are $\beta=1,4$ ensembles (the orthogonal and symplectic ones) which produce Feynman graphs, each one may be
embedded either in orientable surface (in case it is a ribbon graph) or to non-orientable one (in case it is a ribbon
graph with crosscaps).
It was shown in \cite{L1} that partition functions of these ensembles are BKP tau functions.
However the perturbation series written as series of the Schur functions (see \cite{OST-I}) are not
series we need\footnote{Recently appeared the paper \cite{Carrel} where the graph counting of $\beta=1,2$ ensembles was
studied in details.}. To get  the projective Hurwitz numbers we suggest another matrix integrals. These integrals
in the integration measure
contain the simplest BKP tau function $\tau^{B}_1$ which is widely used in our paper, see
(\ref{Laplace=sum-Schurs}),(\ref{r=1}),(\ref{tau-via-vertex-t}),
(\ref{Example-vertex}) and (\ref{vac-tau-BKP'}) below.
Such integral of a BKP tau function may be BKP tau function again but may be not a tau function, the last case occurs
when the integral generate Hurwitz numbers with arbitrary profiles in two and more branch points. We also point out
that the multiple usage of the BKP tau function $\tau_1^{\rm B}$ to deform integration measures of matrix integrals allows to get Hurwitz numbers
related to arbitrary Euler characteristic of the base surface.

To end the introduction let us note that if in (\ref{BKP-hyp-tau})  we take $r$ as in (\ref{r-MirMorNat}) and choose
$\bpow=(1,0,0,\dots)$,
then  (\ref{BKP-hyp-tau}) is a discrete version of the partition function
of the orthogonal ensemble of random matrices:
\be\label{discrete-orthogonal}
 \tau^{\rm B} =\frac{1}{g(n)N!} \sum_{h_1,\dots,h_N\ge 0} \prod_{i<j}|h_i-h_j| \prod_{i=1}^{N} \frac{e^{V(p^*,h_i)}}{h_i!},
\ee
\be\label{V}
V(p^*,x):=\sum_{m>0}\,\frac 1m x^m p^*_m
\ee
where as we shall see, the variables $\zeta$ and $\bpow^*$ are related via $V(\bpow^*,x-1)-V(\bpow^*,x)=V(\zeta ,x)$.
From \cite{L1} we know that (\ref{discrete-orthogonal}) is the BKP tau function with the variables $\bpow^*$
playing the role of  BKP higher times. The factor $g(n)$ is written down in the Appendix \ref{fermionic-appendix}.

In a similar way we may obtain a discrete analogue of the circular $\beta=1$ ensemble
choosing (\ref{r-new-trig}), see Remark \ref{circular-beta=1} in
Section \ref{BKP-tau-function} which proves that for a certain specification of $\bpow$ the series
(\ref{BKP-hyp-tau}) is a BKP tau function with respect to the variables $\xi$. 

The relation (\ref{discrete-orthogonal}) may be interesting  because
$\beta=1$ ensembles generates Mobius graphs related to $n$-gulations  of non-orientable surfaces, see \cite{Mulase}
and references therein.

Now we shall study the above in detail. This paper is a development of our preprint \cite{NO-2014}.

\section{Hurwitz numbers\label{definitions-section}}

\subsection{Definitions and examples}

For a partition $\Delta$ of a number $d=|\Delta|$ denote by $\ell(\Delta)$ the number of the non-vanishing parts.
For the Young diagram corresponding to $\Delta$, the number $|\Delta|$ is the weight  of the diagram  and $\ell(\Delta)$
is the number of rows. Denote by $(d_1,\dots,d_{\ell})$ the Young diagram with rows of length $d_1,\dots,d_{\ell}$ and
corresponding partition of $d=\sum d_i$. We need the notion of the colength of a partition $\Delta$ which is
$\ell^*(\Delta):=|\Delta|-\ell(\Delta)$.

Let us consider a connected compact surface without boundary $\Omega$ and a branched covering $f:\Sigma\rightarrow\Omega$
by a connected or non-connected surface $\Sigma$. We will consider a covering $f$ of the degree $d$. It means that the
preimage $f^{-1}(z)$ consists of $d$ points $z\in\Omega$ except some finite number of points. This points are called
\textit{critical values of $f$}.

Consider the preimage $f^{-1}(z)=\{p_1,\dots,p_{\ell}\}$ of $z\in\Omega$. Denote by $d_i$ the degree of $f$ at $p_i$. It
means that in the neighborhood of $p_i$ the function $f$ is homeomorphic to $x\mapsto x^{d_i}$. The set $(d_1\dots,d_{\ell})$
is the partition of $d$, that is called \textit{topological type of $z$}.

Fix now points $z_1,\dots,z_{\textsc{f}}$ and partitions $\Delta^{(1)},\dots,\Delta^{(\textsc{f})}$ of $d$. Denote by
\[\widetilde{C}_{\Omega (z_1\dots,z_{\textsc{f}})} (d;\Delta^{(1)},\dots,\Delta^{(\textsc{f})})\]
the set of all branched covering $f:\Sigma\rightarrow\Omega$ with critical points $z_1,\dots,z_{\textsc{f}}$ of topological
types  $\Delta^{(1)},\dots,\Delta^{(\textsc{f})}$.

Coverings $f_1:\Sigma_1\rightarrow\Omega$ and $f_2:\Sigma_2\rightarrow\Omega$ are called isomorphic if there exists an
homeomorphism $\varphi:\Sigma_1\rightarrow\Sigma_2$ such that $f_1=f_2\varphi$. Denote by $\texttt{Aut}(f)$  the group of
automorphisms of the covering $f$. Isomorphic coverings have isomorphic groups of automorphisms of degree $|\texttt{Aut}(f)|$.

Consider now the set $C_{\Omega (z_1\dots,z_{\textsc{f}})} (d;\Delta^{(1)},\dots,\Delta^{(\textsc{f})})$ of isomorphic classes
in $\widetilde{C}_{\Omega (z_1\dots,z_{\textsc{f}})} (d;\Delta^{(1)},\dots,\Delta^{(\textsc{f})})$. This is a finite set.
The sum
\[
H^{\textsc{e},\textsc{f}}(d;\Delta^{(1)},\dots,\Delta^{(\textsc{f})})=
\sum\limits_{f\in C_{\Omega (z_1\dots,z_{\textsc{f}})}(d;\Delta^{(1)},\dots,
\Delta^{(\textsc{f})})}\frac{1} {|\texttt{Aut}(f)|}\quad,
\]
don't depend on the location of the points $z_1\dots,z_{\textsc{f}}$ and is called \textit{Hurwitz number}.
Here $\textsc{f}$ denotes the number of the branch points, and $\textsc{e}$ is the Euler characteristic of the base surface.

\vspace{1ex}
{\bf Example}.
Let $f:\Sigma\rightarrow\mathbb{RP}^2$ be a covering without critical points.
Then, if $\Sigma$ is connected, then $\Sigma=\mathbb{RP}^2$,
$\deg f=1$\quad or $\Sigma=S^2$, $\deg f=2$. Therefore if $d=3$, then
$\Sigma=\mathbb{RP}^2\coprod\mathbb{RP}^2\coprod\mathbb{RP}^2$ or $\Sigma=\mathbb{RP}^2\coprod S^2$.
Thus $H^{1,0}(3)=\frac{1}{3!}+\frac{1}{2!}=\frac{2}{3}$.

\vspace{1ex}

The Hurwitz numbers arise in different fields of mathematics: from algebraic geometry to integrable systems. They are well
studied for orientable $\Omega$. In this case the Hurwitz number coincides with the weighted number of holomorphic branched
coverings of a Riemann surface $\Omega$ by other Riemann surfaces, having critical points $z_1,\dots,z_\textsc{f}\in\Omega$ of
the topological types $\Delta^{(1)},\dots,\Delta^{(\textsc{f})}$ respectively. The well known isomorphism between Riemann
surfaces and complex algebraic curves gives the interpretation of the Hurwitz numbers as the numbers of morphisms of
complex algebraic curves.

Similarly, the Hurwitz number for a non-orientable surface $\Omega$ coincides with the weighted number of the dianalytic
branched coverings of the Klein surface without boundary by another Klein surface and coincides with the weighted number
of morphisms of real algebraic curves without real points \cite{AG,N90,N2004}. An extension of the theory to all Klein surfaces
and all real algebraic curves leads to Hurwitz numbers for surfaces
with boundaries may be found in \cite{AN,N}.

\vspace{2ex}

The Hurwitz numbers have a purely algebraic description.  Any branched covering $f:\Sigma\rightarrow\Omega$ with
critical points $z_1,\dots,z_\textsc{f}\in\Omega$ generates a homomorphism $
\phi: \pi_1(u,\Omega\setminus\{z_1,\dots z_F\})\rightarrow S_{\Gamma}$,
where $u$ is a point in $\Omega$,
to the group of permutations of the set $\Gamma=f^{-1}(u)$ by the monodromy along contours of
$\pi_1(u,\Omega\setminus\{z_1,\dots z_F\})$.  Moreover, if $l_i\in\pi_1(u,\Omega\setminus\{z_1,\dots z_F\})$ is a contour
around $z_i$, then the cyclic type of the permutation $\phi(l_i)$ is $\Delta^{(i)}$. Denote by
$$\texttt{Hom}_\Omega(d;\Delta^{(1)},\dots,\Delta^{(\textsc{f})}),$$
the group of all homomorphisms $\phi: \pi_1(u,\Omega\setminus\{z_1,\dots z_F\})\rightarrow S_{\Gamma}\cong S_d$ with this
property. Isomorphic coverings generate elements of $
\texttt{Hom}_\Omega(d;\Delta^{(1)},\dots,\Delta^{(\textsc{f})})$ conjugated by $S_d$. Thus
we construct the one-to-one correspondence between $
C_{\Omega (z_1\dots,z_{\textsc{f}})} (d;\Delta^{(1)},\dots,\Delta^{(\textsc{f})})$ and the conjugated classes
of $\texttt{Hom}_\Omega(d;\Delta^{(1)},\dots,\Delta^{(\textsc{f})})$.

\vspace{1ex}

Consider the last set in more details. Any $s\in S_d$ generates the interior automorphism $I_s(g)=sgs^{-1}$ of $S_d$. Therefore
$S_d$ acts on $\texttt{Hom}_\Omega(d;\Delta^{(1)},\dots,\Delta^{(\textsc{f})})$ by $s(h)=I_s h$. The orbit of this action of
$I=\{I_s\}$ corresponds to an equivalent class of coverings. Moreover, the group $\texttt{A}= \{s\in S_d|s(h)=h\}$ is
 isomorphic to the group
$\texttt{Aut}(f)$, there the covering $f$ corresponds to the homomorphism $h$.

Consider the splitting
$\texttt{Hom}_\Omega(d;\Delta^{(1)},\dots,\Delta^{(\textsc{f})})= \bigcup\limits_{i=1}^rH_i$ on obits by $I$. Then the
cardinality $|H_i|$ is $\frac{d!}{|\texttt{A}(h_i)|}= \frac{d!}{|\texttt{Aut}(f_i)|}$, where $h_i\in H_i$. On the other
hand, the orbits $H_i$ are in the one-to-one correspondence with the classes of the coverings.
Therefore $\frac{1}{d!}|\texttt{Hom}_\Omega(d;\Delta^{(1)}, \dots,\Delta^{(\textsc{f})})| =
\frac{1}{d!}\sum\limits_{i=1}^r |H_i|= \sum\limits_{i=1}^r \frac{1}{|\texttt{Aut}(f_i)|}$ is the
Hurwitz number $H_\Omega(d; \Delta^{(1)}\,\dots,\Delta^{(\textsc{f})})$.

\vspace{2ex}

Find now $|\texttt{Hom}_\Omega(d;\Delta^{(1)},\dots,\Delta^{(\textsc{f})})|$ in terms of the characters of $S_d$. Recall, that
cyclic type  of $s\in S_d$ is cardinalities $\Delta=(d_1,\dots,d_{\ell})$ of subsets, on which the permutation $s$ split
the set $\{1,\dots,d\}$. Any partition $\Delta$ of $d$ generates the set $C_{\Delta}\subset S_d$,  consisting of permutations
of cyclic type $\Delta$. The cardinality of $C_\Delta$ is equal to
\be\label{C-Delta,z-Delta}
|C_\Delta| \,=\,\frac{|\Delta|!}{z_\Delta}\,,\qquad
z_\Delta\,=\,\prod_{i=1}^\infty \,i^{m_i}\,m_i!
\ee
where $m_i$ denotes the number of parts equal to $i$ of the partition $\Delta$ (then the partition $\Delta$ is often
denoted by $(1^{m_1}2^{m_2}\cdots)$).

Moreover, if $s_1,s_2\in C_{\Delta}$, then $\chi(s_1)=\chi(s_2)$ for any complex character $\chi$ of $S_d$. Thus we can
define $\chi(\Delta)$ for a partition $\Delta$, as $\chi(\Delta)=\chi(s)$ for $s\in C_\Delta$.

The Mednykh-Pozdnyakova-Gareth A. Jones formula \cite{M1,M2,ZL,GARETH.A.JONES} says that
\[ \big|\texttt{Hom}_\Omega(d;\Delta^{(1)},\dots, \Delta^{(\textsc{f})})\big| = \,
d!\sum_{\lambda} \,
\left(\frac{{\rm dim}\lambda}{d!}\right)^{\textsc{e}}\,
\prod_{i=1}^\textsc{f} \,\,|C_{\Delta^{(i)}}|\,\,
\frac{\chi(\Delta^{(i)})}
{{\rm dim}\lambda} \,,
\]
where $\textsc{e}=\textsc{e}(\Omega)$ is the Euler characteristic of $\Omega$ and $\chi$ ranges over the irreducible
complex characters of
$S_d$, associated with Young diagrams of weight $d$. Thus we obtain (\ref{Hurwitz-counting}).

In particulary for the projective plane $\mathbb{RP}^2$ we get the relation (\ref{Hurwitz-counting}) where
$\textsc{e}=1$.

\vspace{1ex}

{\bf Example}. Let $\textsc{e}=1$, $\textsc{f}=0$, $d=3$.
Then, $ H^{1,0}(3) = \sum_{|\lambda|=3}
\frac{{\rm dim}\lambda}{d!}=\frac{4}{6}=\frac{2}{3}$.
In general for the unbranched covering  of $\mathbb{RP}^2$ we get the following generating function
(compare to (\ref{r=1}) below)
\be\label{unbranched}
e^{\frac {c^2}{2} + c}\,=\,\sum_{d\ge 0}\, c^d H^{1,0}(d)
\ee
The exponent reflects the fact that the connected unbranched covers of the projective plane may consist of either
the projective plane (the term $c$: 1-fold cover) or the Riemann sphere (2-fold cover, the term $\frac{c^2}{2}$ where $2$ in the
denominator is the order of the automorphisms of the covering by the sphere).

At the end we write down purely combinatorial definition of the projective Hurwitz numbers \cite{M2}, \cite{GARETH.A.JONES}.
Let us consider the symmetric group $S_d$ and  the equation
\be\label{equation-in-S-d}
R^2 X_1 \cdots X_{\textsc{f}} = 1,\quad R,X_i \in S_d,\quad X_i \in C_{\Delta^{(i)}},\, i=1,\dots,\textsc{f}
\ee
where $C_{\Delta^{(i)}},\, i=1,\dots,\textsc{f}$ are the cycle classes of a given set of partitions
$\Delta^{(i)},\,i=1,\dots,\textsc{f}$ of a given weight $d$. Then
$H^{1,\textsc{f}}(d;\Delta^{(1)},\dots,\Delta^{(\textsc{f})})$ is the number of solutions to (\ref{equation-in-S-d}) divided
over $d!$. Say, for unbranched 3-fold covering we get 4 solutions to $R^2=1$ in $S_3$: the unity element and three transpositions.
Thus $H^{1,0}=4:3!$ as we obtained in the Examples above. The number of solutions to $R^2=1$ in $S_d$ is given by
(\ref{unbranched}).

\subsection{Remarks on Mednykh-Pozdnyakova-Gareth A. Jones character formula\label{Remarks-on-Mednykh-subsection}}

Let us start from the following preliminary
\br\label{Natanzon}

It follows from R.Dijkgraaf paper \cite{Dijkgraaf} that the Hurwitz numbers for closed orientable surfaces form 2D topological field theory.
An extension of this result to the case of Klein surfaces (thus to orientable and non-orientable surfaces) was found in \cite{AN},
theorem 5.2. (see also Corollary 3.2 in \cite{AN2008}) On the other hand Mednykh-Pozdnyakova-Gareth A. Jones formula gives the description of
the Hurwitz numbers in terms of characters of the symmetric groups. In this Subsection in fact we give the interpretation  of axioms of
the Klein topological field theory \cite{AN} for Hurwitz numbers in terms of characters of symmetric groups, this approach is different
from \cite{AN}.

\er

(A) Let us present the following simple statement
\bl
\label{Hurwitz-down-Lemma}
\be\label{Hurwitz=Hurwirz-Hurwitz}
H^{\textsc{e}+\textsc{e}_1,\textsc{f}+\textsc{f}_1}(d;\Delta^{(1)},\dots,\Delta^{(\textsc{f}+\textsc{f}_1)})=
\sum_{\Delta}\frac{d!}{|C_\Delta|}
H^{\textsc{e}+1,\textsc{f}+1}(d;\Delta^{(1)},\dots,\Delta^{(\textsc{f})},\Delta)
H^{\textsc{e}_1+1,\textsc{f}_1+1}(d;\Delta,\Delta^{(\textsc{f}+1)},\dots,\Delta^{(\textsc{f}_1)})
\ee
In particular
\be\label{Hurwitz-down}
H^{\textsc{e}-1,\textsc{f}}(d;\Delta^{(1)},\dots,\Delta^{(\textsc{f})})=
\sum_{\Delta}\,
H^{\textsc{e},\textsc{f}+1}(d;\Delta^{(1)},\dots,\Delta^{(\textsc{f})},\Delta) \chi(\Delta)
\ee
where $\chi(\Delta)=\frac{d!H^{1,1}(d;\Delta)}{|C_\Delta|}$ are rational numbers explicitly defined by a partition $\Delta$ as follows
\be\label{delta(Delta)}
\chi(\Delta)=\sum_{\lambda \atop |\lambda|=|\Delta|} \chi_\lambda(\Delta)=\left[
\prod_{i>0,\,{\rm even}}e^{\frac {i}{2}\frac{\partial^2}{\partial p_i^2}}\cdot p_i^{m_i}
\prod_{i>0,\,{\rm odd}}e^{\frac {i}{2}\frac{\partial^2}{\partial p_i^2}+
\frac{\partial}{\partial p_i}}\cdot p_i^{m_i}
\right]_{\bpow =0}
\ee
where $\chi_\lambda(\Delta)$ is the character of the representation $\lambda$ of the symmetric group
$S_d$, $d=|\lambda|$, evaluated on the cycle class $\Delta=(1^{m_1}2^{m_2}\cdots)$.

\el
As a Corollary we get that the Hurwitz numbers of the projective plane may be obtained from
the Hurwitz numbers of the Riemann sphere while the Hurwitz numbers of the torus and of the Klein bottle (see (\ref{E=0}))
may be obtained from Hurwitz numbers of the projective plane.

First we prove the second equality in (\ref{delta(Delta)}). It
follows from the relations
\be\label{Laplace=sum-Schurs}
e^{\sum_{i>0}\frac {i}{2}\frac{\partial^2}{\partial p_i^2}+
\sum_{i>0,\,{\rm odd}}\frac{\partial}{\partial p_i}}=\sum_{\lambda} s_\lambda({\tilde\partial})
\ee
\be\label{Schur-orthogonality}
 \left[s_\lambda({\tilde\partial})\cdot s_\mu(\bpow)   \right]_{\bpow=0}=\delta_{\lambda,\mu}\,,
\qquad
p_1^{m_1}p_2^{m_2}\cdots =:p_\Delta =\sum_{\lambda} \chi_\lambda(\Delta) s_\lambda(\bpow)
\ee
where $s_\lambda({\tilde\partial})$ is $s_\lambda(\bpow)$ where each $p_i$ is replaced by
$i\frac{\partial}{\partial p_i}$. The heat operator in the left hand side of (\ref{Laplace=sum-Schurs}) plays an important role.
The relations (\ref{Schur-orthogonality})
may be found in \cite{Mac}. The relation (\ref{Laplace=sum-Schurs}) is derived from the known relation
\be\label{sum-Schurs}
 \sum_\lambda s_\lambda(\bpow({\bf x}))=\prod_{i<j}\frac{1}{1-x_ix_j}\prod_{i}\frac{1}{1-x_i},
 \quad p_m({\bf x}):=\sum_i x_i^m
\ee
which also may be found in \cite{Mac}.

Equality (\ref{Hurwitz=Hurwirz-Hurwitz}) follows from the orthogonality relation for characters:
$\sum_\Delta |C_\Delta| \chi_\lambda(\Delta)\chi_\mu(\Delta) = d!\delta_{\mu,\lambda}$ where $|\mu|=|\lambda|=|\Delta|=d$
which yields $\sum_\Delta \varphi_\lambda(\Delta)\chi(\Delta)=d!\left({\rm dim}(\lambda)\right)^{-1} $.  Then the formula
(\ref{Hurwitz-counting})  gives (\ref{Hurwitz=Hurwirz-Hurwitz}).

In (\ref{2KP-to-BKP}) below we shall see that the heat operator which enters formula (\ref{delta(Delta)}) also links solutions of 2KP (TL)
and BKP hierarchies.

(B) Another remark is as follows.

Let us use the so-called Frobenius notation \cite{Mac} for a partition $\lambda$: $\lambda=(\alpha_1,\dots,\alpha_{\kappa}|\beta_1,\dots,\beta_\kappa)$,
$\alpha_1>\cdots >\alpha_\kappa\ge 0$, $\beta_1> \cdots >\beta_\kappa \ge 0$. The integer $\kappa=\kappa(\lambda)$ denotes the length
of the main diagonal of the Young diagram $\lambda$, the length of $\lambda$ is denoted by $\ell(\lambda)$.

\bl\label{character-on-d-cycle-lemma}
The normalized character labeled by $\lambda$ evaluated at the cycle $(d)$ (as usual $d=|\lambda|$) vanishes if $\kappa(\lambda)>1$, moreover
\be\label{character-on-d-cycle}
\varphi_\lambda\left((d)\right) = (-1)^{\ell(\lambda)+1}\left( \frac{d!}{{\rm dim}\lambda}  \right)\frac 1d \delta_{1,\kappa(\lambda)}
\ee
\el
For the proof first we notice that the Schur function of a one-hook partition, say, $(\alpha_i|\beta_j)$ is of form
\[
 s_{(\alpha_i|\beta_j)}(\bpow) = \frac 1d (-1)^{\beta_j}p_{\alpha_i+\beta_j+1} + \cdots
\]
where dots denote terms which do not depend on $p_a$, $a\ge \alpha_i+\beta_j+1$ (this fact may be derived, say, from the Jacobi-Trudi
formula $s_\lambda(\bpow)=\det\, s_{(\lambda_i-i+j)}(\bpow)$). Then from the Giambelli identity
$s_\lambda(\bpow)=\det\,s_{(\alpha_i|\beta_j)}(\bpow)$ it follows that $s_\lambda$ does not depend on $p_a$, $a>\alpha_1+\beta_1+1$. Thus
it does not depend on $p_a$, $a\ge d>\alpha_1+\beta_1 +1$ in case $\kappa(\lambda)>1$. Due to the character map relation it means that
$\varphi_\lambda\left((d)\right)=0$ if $\kappa(\lambda)>1$. In case of a one-hook partition $\lambda=(\alpha_1|\beta_1)$ we have
$\alpha_1+\beta_1+1 =d$ and the character
map formula (\ref{Schur-char-map}) yields
\[
 s_{(\alpha_1|\beta_1)}(\bpow)=\frac{{\rm dim}\lambda}{d!}\left(p_d\varphi_\lambda\left((d)\right) +\cdots \right)
\]
where dots denote terms which does not depend on $p_d$. We compare last two formulae and get (\ref{character-on-d-cycle}).

Relation (\ref{character-on-d-cycle}) allows to equate Hurwitz numbers related to different
Euler characteristics of
base Klein surfaces if in both
cases there are nonvanishing numbers of ramification profiles $(d)$. Namely,
 the Mednykh-Pozdnyakova-Gareth A. Jones character formula (\ref{Hurwitz-counting}) yields
\bp\label{d-cycle-proposition} For any natural number $\textsl{g}$
\be
H^{\textsc{e}-2\textsl{g}, \textsc{f}+1}\left(d;\Delta^{(1)},\dots,\Delta^{(\textsc{f})},(d)  \right)=
d^{2\textsl{g}}
H^{\textsc{e}, \textsc{f}+2\textsl{g}+1}\left(d;\Delta^{(1)},\dots,\Delta^{(\textsc{f})},(d),
\underbrace{(d),\dots,(d)}_{2\textsl{g}}  \right)
\ee
\ep
First it was proven by Zagier (for the case of even $\textsc{e}$), see Appendix A in \cite{ZL}. We get it in a different way.

For $d$-fold covers we shall call a branch point \textit{maximally ramified} in case its ramification profile is $(d)$.
\br\label{connected-via-(d)-profile}
Notice that the presence of the profile $(d)$ means that the Hurwitz numbers of the connected and disconnected covering are equal:
$H^{\textsc{e},\textsc{f}}_{\rm connected}(d;(d),\dots)=H^{\textsc{e},\textsc{f}}(d;(d),\dots)$ where dots denote the same set of
ramification profiles.
\er

\br
\label{Zagier}
In the Appendix A by Zagier in \cite{ZL} the polynomial
$\texttt{R}_\Delta(\texttt{q}):=\frac{\prod_{i=1}^{\ell{\Delta}}(1-\texttt{q}^{d_i})}{1-\texttt{q}}=
:\sum_r (-1)^r\texttt{q}^r \chi_r(\Delta) $
was studied. It was shown that $\chi_r $ ($0\le r \le d-1 $) is the character of the irreducible representation of $S_d$ given
by $\chi_r(g)=\tr (g,\pi_r)$, $g\in S_d$,
$\pi_r={\wedge}^r({\rm St}_d) $. Here ${\rm St}_d$ is the vector space $\{ (x_1,\dots,x_d)\in \mathbb{C}^d | x_1+\cdots+x_d=0 \}$ and
$S_d$ acts by permutations of the coordinates.
We can show that $\chi_r$ coincides with $\chi_\lambda$ where $\lambda=(d-r|r)$. 
To do it let us consider the Schur function
$s_\lambda(\bpow(\texttt{q},0))$ where $p_m(\texttt{q},0):=1-\texttt{q}^m$ and $\lambda$ is not yet fixed. We get 
$$
s_\lambda(\bpow)=(-1)^{\ell(\lambda)-1} (1-\texttt{q})\texttt{q}^{\ell(\lambda)-1} \delta_{\kappa(\lambda),1}\, ,
$$
see \cite{Mac}. In the last relation
denote $k=\ell(\lambda)-1$, then $\lambda=(d-k|k)$.
On the other hand formula (\ref{Schur-char-map}) says 
$$
s_{(d-k|k)}(\bpow(\texttt{q},0))=(1-\texttt{q})\sum_{\Delta} \frac{|C_\Delta |}{d!} \chi_{(d-k|k)}(\Delta) \texttt{R}_\Delta(\texttt{q})=
(1-\texttt{q})\sum_{r=0}^{d-1}(-\texttt{q})^r \sum_\Delta \frac{|C_\Delta|}{d!}\chi_{(d-k|k)}(\Delta)\chi_r(\Delta)
$$
Compare last relations.
The orthogonality of characters results to $\chi_k = \chi_{(d-k|k)}$. It means that in the presence of a maximally ramified branch 
point the summation range in (\ref{Hurwitz-counting}) is restricted to one-hook partitions $\lambda$.

 Let us also note that the polynomials $\texttt{R}_\Delta$ are
related to ramification weights (\ref{content-t-q}) below.
\er

In what follows we shall see that tau functions generate Hurwitz numbers containing the maximally ramified branch points.

\section{Content products\label{Content products}}

The content product which enters (\ref{BKP-hyp-tau}) may be written in form of generalized Pochhammer symbol
\be
\prod_{(i.j)\in\lambda}r(x+j-i)=\prod_{i=1}^{\ell(\lambda)} r_{\lambda_i}(x-i+1)
\ee
where $r_n(x):=r(x)r(x+1)\cdots r(x+n-1)$, and also in form of a sort of Boltzmann weight
\be\label{r-via-U}
\prod_{(i.j)\in\lambda}r(x+j-i)=e^{-U_\lambda(x)}:=\prod_{i=1}^{\ell(\lambda)} e^{U_{h_i(0)+x}-U_{h_i(\lambda)+x}}
=\prod_{i=1}^{\kappa(\lambda)} e^{U_{\alpha_i+x}-U_{-\beta_i+x}}
\ee
$\alpha_i, \beta_i,\,i=1,\dots,\kappa$ are the Frobenius coordinates of the partition $\lambda$,
$\lambda=(\alpha_1,\dots\alpha_\kappa|\beta_1,\dots,\beta_\kappa)$ ($\kappa=\kappa(\lambda)$ in the length
of the main diagonal of the Young diagram of the partition $\lambda$),
\be
\quad h_i(\lambda):=\lambda_i-i,\quad r(x)=:\exp (U_{x-1}-U_{x})
\ee
The numbers $U_x$ may be fixed by $U_{x_0}=0$ with a chosen $x_0$. In the present paper $U_x$ is chosen either
as $V(\zeta,x)$ (the parametrization I) or as $V(\xi,\texttt{t}^x)+\xi_0\log\texttt{t} $ (the parametrization II).

\subsection{ Parametrization I \label{parametrization-I-subsection}}

Let us consider
the sums of all normalized characters $\varphi_\lambda$  evaluated on partitions $\Delta$ with a given weight
$d$, $d=|\lambda|=|\Delta|$ and a given length $\ell(\Delta)=d-k$:
\be\label{phi}
\phi_k(\lambda)\,:=\,\sum_{\Delta\atop \ell(\Delta)=d-k} \,\varphi_\lambda(\Delta),\quad k=0,\dots,d-1
\ee
\br \label{special-cases}
Let us note that $\phi_0(\lambda)=1$.
There are two other special cases when the sum of normalized characters (\ref{phi}) contains a single term:

(a) $\phi_{1}(\lambda)=\varphi_\lambda(\Gamma),\, \Gamma=(1^{d-2}2)$ (for $d>1$).
$\phi_1(\lambda)=\varphi_\lambda(\Gamma)$ which is related to the \textit{ minimally ramified} profile:
the profile with the colength equal to 1.
This is the profile of the simple branch point, simple branch points are of main interest in many applications \cite{Dijkgraaf}.

(b) $\phi_{d-1}(\lambda)=\varphi_\lambda((d))$ which is related to the cyclic profile which describes the
\textit{ maximally ramified}  profile (this profile plays a specific role, see Proposition \ref{d-cycle-proposition}).

In what follows we shall use sums $\phi_k$ as building blocks to construct weighted sums of the Hurwitz numbers
(see for instance (\ref{S}) below). Then the cases (a),(b) produce not the weighted sums but Hurwitz numbers themselves (see
(\ref{tau-Hurwitz-themselves}) below).
\er

\br\label{Hurwitz-genus-formula} The quantity $d-\ell(\lambda)$ which is used in the definition (\ref{phi}) is called the
\textit{colength} of a partition $\lambda$ and will be denoted by $\ell^*(\lambda)$.
The colength enters the so-called Riemann-Hurwitz formula which relates the Euler characteristic of a base surface,
$\textsc{e}$, to the Euler characteristic of it's $d$-branched cover,  $\textsc{e}'$ as follows
\[
 \textsc{e}'- d\textsc{e}+\sum_{i} \ell^*(\Delta^{(i)})=0
\]
where the sum ranges over all branch points.
\er

Let us introduce
\be\label{degree}
{\rm deg} \, \phi_k(\lambda)= k
\ee
This degree is equal to the colength of ramification profiles in formula (\ref{phi}), and due to Remark \ref{Hurwitz-genus-formula}
it will be important later to define the Euler characteristic of the covering surfaces in the parametrization I cases.

Next we need
\bl\label{lemma-Phi}
The power sums of the contents of all nodes of a Young diagram $\lambda$ may be expressed in terms of the normalized
characters and ratios of the Schur functions, and
are polynomials in the variables $\phi_k,\,k=1,2,\dots$:
\bea\label{H-m}
 \Phi_m(\lambda) &:=& \, \sum_{(i.j)\in\lambda} (j-i)^m\,=
 \\
 \label{r-via-varphi}
&=& \frac{m}{2\pi i}\oint  a^{m} \prod_{k=1}^{m}
\left(1+\sum_{\Delta} \left(
e^{2\pi i\frac km}a\right)^{-\ell^*(\Delta)}\varphi_\lambda(\Delta)\right)\, \frac{da}{a}
\\
\label{r-via-Schurs}
&=&
(-1)^{m+1}m
\frac{1}{2\pi i}\oint  a^{m}
\log\frac{s_\lambda(\bpow(a))}{s_\lambda(\bpow_\infty)}\, \frac{da}{a}
\\
\label{r-via-log-phi}
&=&
(-1)^{m+1}m
\frac{1}{2\pi i}\oint  a^{m}
\log\left(1+\sum_{k=1}^{d-1} a^{-k}\phi_k(\lambda)\right)\, \frac{da}{a}
\\
\label{Phi}
&=& m\sum_{\mu \atop |\mu|=m\,,\mu_1<d}
 (-1)^{\ell^*(\mu)}(\ell(\mu)-1)!  \frac{\phi_\mu(\lambda)}{{\rm Aut}\,\mu}
\eea
where $m\ge 0$, $|\lambda|=|\Delta|$ and $\ell^*(\mu):= |\mu|-\ell(\mu)$
 is the colength of the partition $\mu$.
\be\label{power-sum-mon-phi}
 \phi_\mu(\lambda):=\prod_{i=k}^{d-1}\,\left(\phi_k(\lambda)\right)^{m_k}
 =\prod_{i=1}^{\ell(\mu)}\,\phi_{\mu_i}(\lambda)\,,\quad \mu=(1^{m_1}\dots(d-1)^{m_{d-1
 }}) =(\mu_1,\dots,\mu_{\ell})
 \ee
 In (\ref{r-via-Schurs}) ${\bf p}(a)=(a,a,\dots)$ and ${\bf p}_\infty=(1,0,0,\dots)$, and in (\ref{Phi})
  ${\rm Aut}\,\mu =\prod_{i=1}^{\ell(\mu)} m_i!$, $m_i$ denotes the number of times a part $i$ enters the
partition $\mu=(1^{m_1}2^{m_2}\cdots )$.
\el

As we see from (\ref{Phi})-(\ref{power-sum-mon-phi}) each integer $\Phi_m$ is a quasihomogeneous polynomial in the rational numbers $\phi_k$,
and according to (\ref{degree}) we assign the degree as follows:
\be\label{Phi-deg}
{\rm deg}\, \Phi_m(\lambda)=m.
\ee

Let us write down the first three $\Phi_m(\lambda)$ for $|\lambda|=d\ge 4$ in terms of normalized characters $\varphi_\lambda$
using (\ref{Phi}),(\ref{power-sum-mon-phi}),(\ref{phi}).
We obtain
\bea\label{examples-of-Phi}
\Phi_{1}(\lambda)= \varphi_\lambda(\Gamma),\quad
 \Phi_2(\lambda)= \left(\varphi_\lambda(\Gamma)\right)^2 - 2\varphi_\lambda((1^{d-4}2^2))
 - 2\varphi_\lambda((1^{d-3}3^1)), \quad
 \Phi_3(\lambda)=\left(\varphi_\lambda(\Gamma)\right)^3 -
\\
\nonumber
3\varphi_\lambda(\Gamma)
\left(\varphi_\lambda((1^{d-4}2^2))+\varphi_\lambda((1^{d-3}3^1))  \right)+
3\varphi_\lambda((1^{d-4} 4^1))
+3\varphi_\lambda((1^{d-5} 2^1 3^1))
+3\varphi_\lambda((1^{d-6}2^3))
\eea

Assume $d>2$, $m>2$.  As one can see from (\ref{Phi}),(\ref{power-sum-mon-phi}),(\ref{phi}) each $\Phi_m(\lambda)$ is of form
$\left(\varphi_\lambda(\Gamma)\right)^m+\cdots$
where by dots we denote the contribution of cyclic classes marked by partitions, say $\Delta$, whose length $\ell(\Delta)$
belong either to the interval $[d-2,d-m]$ if $m<d$, or to the interval $[d-2,1]$ if $m\ge d$.

 The proof that the right hand side of (\ref{H-m}) is equal to (\ref{r-via-varphi}) is based on two relations
 \[
  \lim_{n\to \infty} \prod_{k=1}^{m}
\left(1 - n^{-\frac 1m}e^{2\pi i\frac km} x  \right)^n = e^{-x^m}
 \]
and
\be\label{content-a}
\prod_{(i.j)\in\lambda} \left(a+j-i\right) \,=\,
a^{|\lambda|}\left(1+\sum_{\Delta} \varphi_\lambda(\Delta) a^{\ell(\Delta)-|\lambda|} \right)
= a^{|\lambda|}\left(1+\sum_{k=1}^{d-1} \phi_k(\lambda) a^{-k} \right)
\ee
which may be obtained from relations in \cite{Mac}. This relation is important and will be further exploited to get Hurwitz
numbers and weighted sums of Hurwitz numbers.

The proof that the right hand side of (\ref{H-m}) is equal
(\ref{r-via-log-phi}) follows from (\ref{content-a}):
\[
 \prod_{i,j \in \lambda} e^{-\sum_{m>0} \frac 1m (-a)^{-m} (j-i)^m} =
 \prod_{i,j \in\lambda}\left(1+\frac{j-i}{a}  \right)= 1+\sum_{k=1}^{d-1} \phi_k(\lambda) a^{-k}
\]

\br By comparing the first and the last terms in (\ref{content-a}) one can conclude that
\[
\phi_k(\lambda)=0,  \quad {\rm if}\, k>d-\kappa(\lambda)
\]
where $\kappa(\lambda)$ in the length of the main diagonal of the Young diagram of the partition $\lambda$.
Now take $k=d-1$ as in (b) of the Remark \ref{special-cases}. Then it follows that
$\varphi_\lambda\left((d) \right)$ is non-vanishing only for one-hook Young diagrams $\lambda=(d-a,1^a)$, $a=0,1,\dots,d$.

\er

\br \label{vanishing-sums}
It follows from the relation (\ref{content-a}) that the following sum vanishes
\[
 a^d\left(1+\sum_{k=1}^{d-1} \phi_k(\lambda)a^{-k}\right)=0
\]
if $a$ is integer and an addition if $-\lambda_1 < a < \ell(\lambda)$.

\er

\br From (\ref{content-a}) we see that
\be
s_{(1^k)}\left(\Phi(\lambda)\right)=\phi_k(\lambda)
\ee
where $\Phi(\lambda)=\left(\Phi_1(\lambda),\Phi_2(\lambda),\dots \right)$

\er

\bp
\label{Proposition-texttt-H}
Let
\be\label{choice-r-1}
r(\zeta,h;x)=\,\exp V(\zeta,hx)
\ee
where $\zeta$ is the infinite set of parameters  $\zeta=(\zeta_1,\zeta_2,\dots)$ and  $V$ is
defined by (\ref{V}). Then the related content product may be expressed in terms of characters in the
following explicit way
\be
\label{content-textttH}
  \prod_{(i.j)\in\lambda}r(\zeta, h;j-i)       =\,\exp \sum_{m>0} \frac 1m h^m\zeta_m\Phi_m
\ee
\ep

From (\ref{r-via-U}) we get
\be\label{content-product-zeta-p^*}
\prod_{(i.j)\in\lambda}r(\zeta,j-i)\,=\,
\prod_{i=1}^{\kappa(\lambda)}
\,e^{V(\bpow^*,\alpha_i)-V(\bpow^*,-\beta_i-1)}=\prod_{i=1}^{L}\, e^{V(\bpow^*,h_i(\lambda))-V(\bpow^*,0)}
\ee
where $h_i(\lambda)=\lambda_i-i$ and $\alpha_i, \beta_i$ are Frobenius coordinates of the partition $\lambda=(\alpha|\beta)$. Then
the variables $\bpow^*=(p_1^*,p_2^*,\dots)$ are related to the variables
$\zeta$ by the triangle transformation given by
\be\label{xi-p*}
V(\zeta,x)=V(\bpow^*,x-1)-V(\bpow^*,x)
\ee
In particular we get the discrete version of the orthogonal ensemble given by (\ref{discrete-orthogonal}).

 \br\label{AMMN-completed-cycles}
With the help of (\ref{content-product-zeta-p^*}) the Proposition \ref{Proposition-texttt-H} may be related to the
well-known results \cite{Okounkov-Pand-2006},
 \cite{AMMN-2011}
 on Hurwitz numbers and the completed cycles as follows.
 In \cite{{AMMN-2011}}
 the generation function for Hurwitz numbers of covers of $\mathbb{CP}^1$ in form
 \be\label{tau-hyp-TL}
\tau^{\rm TL}(\bpow^{(1)},\bpow^{(2)}|\bpow^*)=  \sum_{\lambda}
e^{\sum_{m>0} \frac 1m p_m^* \textsc{C}_\lambda(m) } s_\lambda(\bpow^{(1)})s_\lambda(\bpow^{(2)})
 \ee
was studied and identified with a specification of the KP hypergeometric tau function \cite{KMMM}, \cite{OS-TMP}.
The exponential prefactor in this KP hypergeometric tau function coincides with the right hand side
of (\ref{content-product-zeta-p^*}).
Then it follows from (\ref{xi-p*}) that
\[
\sum_{i=1}^{\ell(\lambda)} \left((\lambda_i-i)^m-(-i)^m\right) =: \textsc{C}_\lambda(m)=
\sum_{k=1}^{m-1} \frac{(-1)^{m-k}}{(m-k)!} \frac{(m-1)!}{(k-1)!}\Phi_k(\lambda)
\]

 \er

 Further remarks on (\ref{tau-hyp-TL}):

 \br \label{remark-kontsevich}
 (A) Let $\bpow^{(1)}=\bpow^{(2)}=(1,0,0,\dots)$ in (\ref{tau-hyp-TL}).
 Then the variables
 $\bpow^*$ may be identified with the KP higher times  because
 the expression (\ref{tau-hyp-TL}) yields a discrete version of the one-matrix model (the unitary ensemble), quite
 similarly to (\ref{discrete-orthogonal}) which describes a discrete model the orthogonal ensemble. (B) If we choose
 $\bpow^{(1)}=\bpow^{(2)}=\bpow(0,\texttt{t})$ (the notation see in the Introduction) and specify $p_m^*$ we obtain the partition
 functions of the $U(N)$ Chern-Simons model on $S^3$ with the coupling constant $g_s=-\log \texttt{t}$ \cite{Szabo}. (C) Let us take
  $\bpow^{(2)}=\bpow(0,\texttt{t})$ and $p_m^*=0,\,m>2$, then the right hand side of (\ref{tau-hyp-TL})
 generates Marino-Vafa relations for Hodge integrals
 \cite{Zhou} (where $p$, $\lambda\tau$ and $\lambda$ are respectively $\bpow^{(1)}$, $p_2^*$ and $\sqrt{-1}\log \texttt{t}$ in our notations).
 (D) It was first noticed in
 \cite{KMMM} (see also  \cite{OShiota-2004})  that for the choice $\bpow^{(1)}=(1,0,0,\dots)$, $p^{(2)}_m=\sum x_i^m$ the
 series (\ref{tau-hyp-TL}) is a discrete version of the Kontsevich model:
 \[
 \tau^{\rm TL}({\bf x},\bpow^*)=
 \frac {1}{N!} \sum_{h_1,\dots,h_N}  \prod_{i<j}(h_i-h_j)\prod_{i=1}^N \frac{1}{h_i!}e^{V(\bpow^*,h_i)+\textsc{l}_ih_i},\quad
 x_i=e^{\textsc{l}_i}
 \]

 \er

\subsection{Parametrization II.}

If $j-i$ is the content of the node of $\lambda$ the number $\texttt{t}^{j-i}$ is called the \textit{quantum content} of the node.

\bl\label{lemma-T}

The power sums of the quantum contents $\texttt{t}^{j-i}$ of all nodes of the Young diagram $\lambda$ is expressed in terms
of the parts of $\lambda$,
Schur functions and the normalized characters $\varphi_\lambda$ as follows
 \bea\label{T-m}
 T_\lambda(\texttt{t}) &:=& \, \sum_{(i.j)\in\lambda} \texttt{t}^{j-i}
 \\
&=& \label{T-via-h-i}
\sum_{i=1}^{\ell(\lambda)} \texttt{t}^{1-i}\frac{1-\texttt{t}^{\lambda_i}}{1-\texttt{t}}  =
 \frac{\texttt{t}}{\texttt{t}-1} \sum_{i=1}^{\ell(\lambda)} \left(\texttt{t}^{h_i(\lambda)}-\texttt{t}^{h_i(0)}\right)
 \\
 \label{log-der-Schur}
&=&
 p_1\frac{\partial}{\partial p_1}\log s_\lambda(\bpow)|_{\bpow=\bpow(0,\texttt{t})}
 \\
 \label{A-q}
&=&
\frac{d+\sum'_{\Delta}\,\texttt{m}_1(\Delta) A_{\Delta}(\lambda,\texttt{t})}
 {1+\sum'_{\Delta} \,A_{\Delta}(\lambda,\texttt{t})},\quad
 A_{\Delta}(\lambda,\texttt{t})\,=\, \varphi_\lambda(\Delta)\, \frac{
 \left(1-\texttt{t}\right)^d}{
 \prod_{j=1}^{\ell(\Delta)}\left(1-\texttt{t}^{d_j}\right) }
\eea
where $h_i(\lambda)=\lambda_i-i$, $|\lambda|=|\Delta|=d$, $\sum'$ denotes the sum over all partitions
except the partition $(1^d)$.
The partition $\Delta$
is written either as $(d_1,\dots,d_{\ell(\Delta)})$ or as $(1^{\texttt{m}_1}2^{\texttt{m}_2}\cdots)$,
$\texttt{m}_i=\texttt{m}_i(\Delta)$ denotes the number of parts of
$\Delta$ equal to $i$. In formula (\ref{log-der-Schur}) first we take the derivative with respect to $p_1$,
then evaluate the power sum variables $\bpow$ as   $\bpow=\bpow(0,\texttt{t}^m)=(p_1,p_2,\dots)$ where
$p_k=p_k(0,\texttt{t}^m)=(1-\texttt{t}^{km})^{-1}$.

\el
The proof is similar to the previous case but instead of (\ref{content-a}) we use another relation:
\be\label{content-t-q}
\prod_{(i.j)\in\lambda} \frac{1-\texttt{q}\texttt{t}^{j-i}}{1-\tilde{\texttt{q}}\texttt{t}^{j-i}} \,=
\frac{s_\lambda(\bpow(\texttt{q},\texttt{t}))}{s_\lambda(\bpow(\tilde{\texttt{q}},\texttt{t}))}
=\,
\left(\frac{1-\texttt{q}}{1-\tilde{\texttt{q}}}\right)^{|\lambda|}
\frac{1+\sum'_{\Delta} \varphi_\lambda(\Delta) w(\Delta,\texttt{q},\texttt{t})}
{1+\sum'_{\Delta} \varphi_\lambda(\Delta) w(\Delta,\tilde{\texttt{q}},\texttt{t})}
\ee
where $\bpow(\texttt{q},\texttt{t})=(p_1(\texttt{q},\texttt{t}),p_2(\texttt{q},\texttt{t}),\dots)$
\be\label{p(t,q)}
p_m(\texttt{q},\texttt{t})=\frac{1-\texttt{q}^m}{1-\texttt{t}^m}
\ee
and
\be\label{w}
w(\Delta,\texttt{q},\texttt{t})\,=
\,\frac{(1-\texttt{t})^{d}}{(1-\texttt{q})^{d}}\prod_{i=1}^{\ell(\Delta)}
\frac{1-\texttt{q}^{d_i}}{1-\texttt{t}^{d_i}}
\ee
may be called $\texttt{q},\texttt{t}$-ramification weight. We have
$w(\Delta,e^{ah},e^{h})\to a^{\ell^*(\Delta)}$ as $h\to 0$.
Equation (\ref{content-t-q})  can be easily obtained from known relations presented in \cite{Mac}.

For the proof we put $\tilde{\texttt{q}}=0$ and replace $\texttt{q} \to \frac {\texttt{q}}{n}$
and consider the $n$-th power of (\ref{content-t-q}) getting (\ref{A-q}) from the right hand side of (\ref{content-t-q})
where we insert (\ref{w}).
 Then (\ref{log-der-Schur}) follows from  (\ref{A-q}).

 \br
 Apart from relations (\ref{T-m})-(\ref{A-q}) one may also write
 \[
  T_\lambda(\texttt{t}^m)=\frac{1}{2\pi i}\oint \texttt{q}^{-1-m} \log
  \frac{s_\lambda(\bpow(\texttt{q},\texttt{t}))}{s_\lambda(\bpow(0,\texttt{t}))} d\texttt{q}\,,\quad m>0
 \]
 which is the analogue of (\ref{r-via-log-phi}).
\er
\br
We get $\Phi_m(\lambda)=\left(\texttt{t}\frac{\partial}{\partial\texttt{t}} \right)^m \cdot T_\lambda(\texttt{t})|_{\texttt{t}=1}$.
\er

 \bp\label{Proposition-texttt-T} Let
\be\label{choice-r-2}
r(\xi,x|\texttt{t})=\,e^{V(\xi_+,\texttt{t}^x) +\xi_0 x\log \texttt{t} +V(\xi_-,\texttt{t}^{-x})}=\,
e^{\sum_{m\neq 0}\frac{1-\texttt{t}^m}{m\texttt{t}^m}\texttt{p}^*_m \texttt{t}^{mx}+\xi_0 x\log \texttt{t}}
\ee
where $\xi$ is the collection of parameters $\xi_0$ and $\xi_\pm=(\xi_{\pm 1},\xi_{\pm 2},\dots)$, and where
$V$ is defined by (\ref{V}). Then
\be\label{content-xi-T}
\prod_{(i.j)\in\lambda}r(\xi,x+j-i|\texttt{t})
=\,
e^{\xi_0 (\varphi_\lambda(\Gamma)+|\lambda|x)\log\texttt{t}   + \sum_{m\neq 0} \frac 1m \xi_m \texttt{t}^{mx} T_\lambda(\texttt{t}^m)}=
\ee
\be\label{p^*-t^h}
= \prod_{i=1}^{\ell(\lambda)}e^{
\frac{\xi_0\log \texttt{t}}{2} \left(\left(x+h_i(\lambda)\right)^2 + \left(x+h_i(\lambda)\right)
 - \left(x+h_i(0)\right)^2 - \left(x+h_i(0)\right)\right)
 +\sum_{m\neq 0}\frac 1m \texttt{p}^*_m \left(\texttt{t}^{(h_i(\lambda) + x)m}-\texttt{t}^{(h_i(0) + x)m}\right)}
 \ee
where $\texttt{p}^*_m=\xi_m\frac{\texttt{t}^m}{\texttt{t}^m-1}$, $h_i(\lambda)=\lambda_i-i,\, h_i(0)=-i$.

 \ep

\br \label{special-q-t-limits}

 The right hand side of (\ref{content-t-q}) may be obtained as the specification of parameters: $x=0$,
 $\xi_m=0$, $m\le 0$, and
 $\xi_m=\tilde{\texttt{q}}^m-\texttt{q}^m,\,m>0$ in
 (\ref{content-xi-T}).
 Relation (\ref{content-t-q}) is of use to get Hurwitz numbers in special cases. However one needs to explain how he would treat
 the denominator in the right hand side. Among others consider three different ways to do it:

 (A)  Let us fix $\tilde{\texttt{q}}$. Then $w(\Delta,\tilde{\texttt{q}},\texttt{t})$ tends to zero if $\texttt{t}\to 1$ for $\Delta\neq (1^d)$. It allows
 to developed the denominator in the right hand side of (\ref{content-t-q}) and also $T_\lambda(\texttt{t})$ of (\ref{A-q}) in
 Taylor series in the normalized  characters of $S_d$ for $\texttt{t}$ close to 1. The limit $\texttt{t}\to 1$ returns us to the case
 studied in the Subsection \ref{parametrization-I-subsection}.

There are also two different limiting procedures which allows to get rid of the infinite sum arising from
the character expansion of $s_\lambda(\bpow(\tilde{\texttt{q}},\texttt{t}))$
in the denominator of the right hand side of (\ref{content-t-q}) in case the leading term in the denominator
is the term with $\Delta=(d)$:

(B) The second one is based on taking $\tilde{\texttt{q}}$ to be close to 1:
 \bl Let $\epsilon$ be a small parameter. Then
 \be\label{epsilon-t}
 \varphi_\lambda\left((d)\right) \prod_{(i,j)\in\lambda}\frac{1-\texttt{q}\texttt{t}^{j-i}}{1-e^\epsilon\texttt{t}^{j-i}}=
 \frac{1}{\epsilon} \delta_{1,\kappa(\lambda)} \frac{1-\texttt{t}^d}{d}
\frac{s_\lambda(\bpow(\texttt{q},\texttt{t}))}{s_\lambda(\bpow_\infty)} + O(1)
\ee
 where we use the notations of Lemma \ref{character-on-d-cycle-lemma} and
 where $O(1)$ denotes terms of order $\epsilon^{k},\,k=0,1,\dots$.
 \el
 The Lemma follows from (\ref{content-t-q}), from $s_\lambda(\bpow_\infty)=\frac{{\rm dim}\lambda}{d!}$, and from the
 formula (\ref{Schur-char-map}) where power sums are specified by (\ref{p(t,q)}),
 in particular we have
 \be
 s_\lambda(\bpow(e^\epsilon,\texttt{t}))=\epsilon \frac{{\rm dim}\lambda}{d!}\varphi_\lambda\left((d)\right)\frac{d}{1-\texttt{t}^d}
 + o(\epsilon)
 \ee
 (C) The third one is based on choosing $\texttt{t}$ to be close to the root of unity $\texttt{t}_0=e^{\frac{2\pi}{d}\sqrt{-1}}$,
 and is as follows.
 Let $\lambda$ be a one-hook partition: $\kappa(\lambda)=1$. As we know $\varphi_\lambda\left((d)\right)\neq 0$ in this case, see Lemma
 \ref{character-on-d-cycle-lemma}.
It is interesting to see what is going on with (\ref{content-t-q}) in case
$\texttt{t}=e^{\frac{2\pi}{d}\sqrt{-1}}e^\epsilon$, where $\epsilon$ is small,
and $\texttt{q}=\texttt{t}^K$ with $K$ being a positive integer, and $K<d$. In this case $w(\Delta,\texttt{t}_0^K,\texttt{t}_0) $
is a finite number which a polynomial in $\texttt{t}_0$. Let $\tilde{\texttt{q}}$ be fixed and $\tilde{\texttt{q}}^d \neq 1$. Then
 \be
\varphi_\lambda\left((d)\right) \prod_{(i,j)\in\lambda} \frac{1-\texttt{t}^{K+j-i}}{1-\tilde{\texttt{q}}\texttt{t}^{j-i}} =
\varphi_\lambda\left((d)\right)\frac{s_\lambda(\bpow(\texttt{t}^K,\texttt{t}))}{s_\lambda(\bpow(\tilde{\texttt{q}},\texttt{t}))}=
\epsilon \frac{d}{1-\tilde{\texttt{q}}^d} \frac{s_\lambda(\bpow(\texttt{t}_0^K,\texttt{t}_0))}{s_\lambda(\bpow_\infty)}
\delta_{1,\kappa(\lambda)}
+o(\epsilon)
 \ee
 We also get
\[
 T_\lambda(\texttt{t}_0e^\epsilon)=\epsilon \frac{d^2}{\varphi_\lambda\left((d)\right)}+o(\epsilon)\,,
 \qquad
s_\lambda(\bpow(0,\texttt{t}_0e^\epsilon))=\frac{1}{\epsilon d}{\varphi_\lambda\left((d)\right)} +
O(1)
\]

 \er

 Let us note the similarity of relations (\ref{p(t,q)})-(\ref{w})
 to the scalar product of the power sums symmetric functions where the Macdonald's symmetric functions are orthogonal,
 see \cite{Mac}. We have
\br\label{Hall-Littlewood} For $\xi_0=\xi_-=0$.
Let us re-write eq. (\ref{content-xi-T}) as follows
\be\label{content-p^*-T}
\prod_{(i.j)\in\lambda}r(\xi,x+j-i|\texttt{t})
=\,
e^{ \sum_{m > 0} \frac 1m (1-\texttt{t}^{m}) \texttt{p}^*_m \texttt{t}^{mx-m} T_\lambda(\texttt{t}^m)}
\,=\,\sum_{\mu}\, \texttt{t}^{(x-1)|\lambda|}
P_\mu({\bf \texttt{p}}^*;0,\texttt{t}) Q_\mu\left(\texttt{T}_{\lambda,\texttt{t}};0,\texttt{t}\right)
\ee
where $P_\mu$ and $Q_\mu$ are Macdonald polynomials with parameters $\texttt{q}$ and $\texttt{t}$ evaluated at the
$\texttt{q}=0$ (namely, these are Hall-Littlewood polynomials).
Here the notations are the same as in \cite{Mac}, however here $P_\mu$ and $Q_\mu$ are written as functions of power
sums variables
which are ${\bf \texttt{p}}^*=(\texttt{p}^*_1,\texttt{p}^*_2,\texttt{p}^*_3,\dots)$,  for $P_\mu$ and
$\texttt{T}_{\lambda,\texttt{t}}=\left(T_\lambda(\texttt{t}),T_\lambda(\texttt{t}^2), T_\lambda(\texttt{t}^3),\dots \right)$
 for the second Hall-Littlewood polynomial $Q_\mu$. Polynomials $Q_\mu$ may be also viewed as the symmetric functions of $d$
  quantum contents $\texttt{t}^{j-i},\,(i,j)\in \lambda$.
We remind \cite{Mac} that the scalar products of power sums and of the Macdonald polynomials with the parameters
$\texttt{q}$ and $\texttt{t}$ may be written as
\[
 <p_\lambda,p_\mu>=z_\mu \prod_{i=1}^{\ell(\mu)}
\frac{1-\texttt{q}^{\mu_i}}{1-\texttt{t}^{\mu_i}} \delta_{\mu,\lambda}\,,\quad <P_\lambda,Q_\mu >=\delta_{\mu,\lambda}
\]
The number $z_\mu$ is defined by (\ref{C-Delta,z-Delta}) below.
The origin of the appearance of the Hall-Littlewood polynomials is not clear.
\er

\section{Weighted sums of Hurwitz numbers \label{Hurwitz-numbers-section}}

Below we will consider
combinations of normalized
characters written as follows
\[
 \sum_{|\lambda|=d\atop \ell(\lambda)\le N} (*)\varphi_\lambda(\Delta) \frac{{\rm dim}\lambda}{d!}
\]
where (*) denotes a chosen (polynomial or not polynomial) function in many variables where the role of variables play
the normalized
characters $\varphi_\lambda$ evaluated at all possible different partitions of the number $d$.
According to (\ref{Hurwitz-counting}) in case $d\le N$ this sum is a weighted sum of the projective Hurwitz numbers.
However the parameter $N$ is an arbitrary integer and may be chosen large enough, so in this work we will not care about this
unequality. The point that the sums below
may obtained as specifications of $\textsc{H}_r(d,\Delta)$ of (\ref{textsc-H}) resulting from the choice of $r$ in either (\ref{r-MirMorNat})
or (\ref{r-new-trig}). Different examples
of specifications presented also in Section \ref{Examples-tau-subsection}.

Weighted sums below may be compared to the weighted sums studied in \cite{HO-2014}, \cite{HarnadMathieu-sept-2014}
where statistics of the $\mathbb{CP}^1$ Hurwitz numbers compatible with the property of the integrability of the related
generating series was studied. Let us notice that
though we can not choose functions (*) in an arbitrary way,
there are infinitely many ways to choose them,
we are interested in those which are related to BKP tau functions in a natural way.

The factor (*) appears due to the content product in the formula for hypergeometric tau functions.

Weighted sums below are labeled by a given partition $\mu=(\mu_1,\mu_2,\dots)$.
Our examples are as follows.

\subsection{Parametrization I}
In case (\ref{r-MirMorNat}) we will weight Hurwitz with the help of symmetric functions of contents
viewed as functions of the power sums variable, the role of power sums play $(\Phi_1(\lambda),\Phi_2(\lambda),\dots)$
defined by (\ref{Phi}) and $(\phi_1(\lambda),\phi_2(\lambda),\dots)$
defined by (\ref{phi}).

(a)  Hurwitz numbers weighted by power sums monomials built of $(\Phi_1(\lambda),\Phi_2(\lambda),\dots)$,
$\Phi_\mu(\lambda):=\prod_{i=1}^{\ell(\mu)} \Phi_{\mu_i}(\lambda)$:
\be\label{texttt-H-mu}
\texttt{C}_{\mu}(d;\Delta)\,:=\,\sum_{\lambda\atop |\lambda|=d}\,
\Phi_\mu(\lambda)\, \varphi_\lambda(\Delta)\frac{{\rm dim}\lambda}{d!}
\ee
This is a linear combination of Hurwitz numbers of (both connected and disconnected) $d$-fold covers with
the profile $\Delta$ at $\infty$ and $\ell(\mu)$ different branch points, and Euler characteristic
of the covers is $\textsc{E}'=\ell(\Delta)-d-|\mu|$. This follows from the Hurwitz formula for a $d$-fold covering
$\textsc{E}'-Ed=\sum_{i}(\ell(\Delta_i)-d)$ where the sum range over all branch points, where $\textsc{E}'$ and
$\textsc{E}$ are Euler characteristic respectively of the cover and of the base.

In case we choose $\mu= (1^b)$, the integer $\texttt{C}_{\mu}(d;\Delta)$ counts the number of branched non-equivalent coverings
of the projective plane with a given ramification profile at some point and $b$ simple branch points
\be\label{BKP-Okounkov-case-Hurwitz}
 \texttt{C}_{(1^b)}(\Delta)= H_{\mathbb{RP}^2}(d;
 \underbrace{\Gamma,\dots,\Gamma }_{b} ,\Delta) ,\qquad |\Gamma|=|\Delta|=d
\ee
For
$\mu=(1^b 2)$ by (\ref{examples-of-Phi}) we obtain
\[
 \texttt{C}_{(1^b2)}(\Delta)=-H_{\mathbb{RP}^2}(d;
 \underbrace{\Gamma,\dots,\Gamma }_{b+2}
 ,\Delta)+2H_{\mathbb{RP}^2}(d;\underbrace{\Gamma,\dots,\Gamma }_{b},(1^{d-4}2^2),\Delta)+
 2H_{\mathbb{RP}^2}(d;\underbrace{\Gamma,\dots,\Gamma }_{b},(1^{d-3}3^1),\Delta)
\]

(b) Hurwitz numbers weighted by Jack polynomials. In our case the Jack polynomials are homogeneous symmetric polynomials in $d$ variables
which are integers - contents of all nodes of $\lambda$.
At the same time the Jack polynomials may be re-written as the (quasihomogeneous) polynomials in power sum variables -
in the integers
$(\Phi_1(\lambda),\Phi_2(\lambda),\dots)$ (which in turn are also quasihomogeneous in the variables $\phi_k(\lambda)$ according to (\ref{Phi})).
The last fact allows to use the content product to define the weighted Hurwitz numbers as follows:
\be\label{texttt-J}
\texttt{J}^{(\alpha)}_{\mu}(d;\Delta)\, :=\,\sum_{\lambda\atop |\lambda|=d}\,
Q^{(\alpha)}_\mu(\Phi(\lambda))\, \varphi_\lambda(\Delta)\frac{{\rm dim}\lambda}{d!}
\ee
where $Q^{(\alpha)}_\lambda$ is the (dual) Jack polynomial in notations of Ch VI, sec 10 of \cite{Mac}.

The Euler characteristic of the cover is $\textsc{e}'= \ell(\Delta)-d-|\mu|$ similar to the previous example.

(c) Perhaps the most important example is the sum of Hurwitz numbers which may be called
projective Goulden-Jackson Hurwitz number (compare to \cite{Goulden-Jackson-2008})
\be\label{S}
\texttt{S}_\mu(d;\Delta) \, :=\,
\sum_\lambda\, \frac{{\rm dim}\lambda}{d!} \varphi_\lambda(\Delta)
\prod_{s=1}^k\phi_{\mu_s}(\lambda)\,
 =\sum_{\Delta^1,\dots,\Delta^s\atop \ell^*(\Delta^{s})=\mu_s\,,\, s=1,\dots,k}\,
 H_{\mathbb{RP}^2}(d; \Delta^{1},\dots,\Delta^{k},\Delta)
\ee
which is the sum of the Hurwitz numbers of all $d$-branched covers of $\mathbb{RP}^2$ with $k+1$ ramification profiles
given by an arbitrary partition $\Delta$ and  partitions $\Delta^s\,,\,s=1,\dots,k$ whose lengths are given numbers:
$\ell(\Delta^s)=d-\mu_s$.  The Euler characteristic of the cover is $\textsc{e}'=\ell(\Delta)-d-|\mu|$.
In (\ref{S}) $\phi_\mu(\lambda):=\prod_{i=1}^{\ell(\mu)} \phi_{\mu_i}(\lambda)$ where $\phi_i$ were introduced in (\ref{phi}).
Actually each weighted sum of Hurwitz numbers obtained from BKP is a linear combination of (\ref{S}).

\br\label{simple+maximally-ramified} Let us select the case where the sum $S_\mu$ reduces to a single term, thus it is not the sum
of Hurwitz numbers but a Hurwitz number itself. We occurs if we choose the partition $\mu$ to be $\mu(b,m):=(1^{b}(d-1)^{m})$. We get
\be
\texttt{S}_{\mu(b,m)}(d,\Delta)=H^{1,b+m+1}(d;\Delta,\underbrace{ \Gamma,\dots,\Gamma}_b,\underbrace{(d),\dots,(d)}_m)
\ee
which counts $d$-fold covers of $\mathbb{RP}^2$ with the following set if ramification profiles: an arbitrary profile $\Delta$, say,
over $0$, and $b$ simple branch points and also $m$ maximally ramified profiles. In case $m>0$ this Hurwitz number coincides with the Hurwitz
numbers of connected covers. It results from Remarks \ref{connected-via-(d)-profile}, \ref{special-cases}.
\er

\subsection{Parametrisation II}
In case (\ref{r-MirMorNat}) we will weight Hurwitz with the help of symmetric functions
viewed as functions of the power sums variable, the role of the set of power sums play the set
$(T_\lambda(\texttt{t}),T_\lambda(\texttt{t}^2),\dots)$
defined by (\ref{T-m}). We denote this set by $\texttt{T}_{\lambda,\texttt{t}}$. In this case weighted sums contain Hurwitz numbers for covers with different Euler characteristic therefore
there is no sense to introduce the analogue of the constant $h$.

 In the examples below the prefactor (*) is not a polynomial function of $\varphi_\lambda$.
For a given  partition $\mu$ we introduce $\texttt{t}$-dependent sums

(d) Hurwitz numbers weighted by the power sums monomials  built of $(T_\lambda(\texttt{t}),T_\lambda(\texttt{t}^2),\dots)$
 \be\label{texttt-T-mu}
 \texttt{K}_{\mu}(d;\Delta|\texttt{t})\,:=\,\sum_{\lambda\atop |\lambda|=d}\,
T_\lambda(\mu|\texttt{t})\, \varphi_\lambda(\Delta)\,\frac{{\rm dim}\lambda}{d!}
,\qquad |\Delta|=d
\ee
where  $T_\lambda(\mu|\texttt{t})=
\prod_{i=1}^{\ell(\mu)} T_\lambda(\texttt{t}^{\mu_i})$ and $T_\lambda(\texttt{t}^{\mu_i})$ are defined by (\ref{T-m}).

(e) $\texttt{t}$-dependent sums weighted by Jack polynomials
\be\label{texttt-J-alpha-t-mu}
 \texttt{J}^{(\alpha)}_{\mu}(d;\Delta|\texttt{t})\,:=
 \sum_{\lambda\atop |\lambda|=d}\,Q^{(\alpha)}_\mu(\texttt{T}_{\lambda,\texttt{t}})
 \, \varphi_\lambda(\Delta)\,\frac{{\rm dim}\lambda}{d!}
,\qquad |\Delta|=d
\ee
where $Q_\mu^{(\alpha)}$ is the Jack polynomial. It may be viewed either as the homogeneous symmetric function of the
$d$ variables - quantum contents of the diagram $\lambda$, or, alternatively, as the quasihomogeneous
functions of power sums variables $\texttt{T}_{\lambda,\texttt{t}} =
\left(T_\lambda(\texttt{t}), T_\lambda(\texttt{t}^2), T_\lambda(\texttt{t}^3),\dots \right)$ expressed in terms of
$S_d$ characters via Lemma \ref{lemma-T} (see also Remark \ref{Hall-Littlewood}).

(f) sums weighted by Macdonald polynomials
\be\label{texttt-M-mu}
 \texttt{M}^{\texttt{q},\texttt{t}}_{\mu}(d;\Delta)\,:=
 \sum_{\lambda\atop |\lambda|=d}\,Q^{\texttt{q},\texttt{t}}_\mu(\texttt{T}_{\lambda,\texttt{t}})
 \, \varphi_\lambda(\Delta)\,\frac{{\rm dim}\lambda}{d!}
,\qquad |\Delta|=d
\ee

where $Q_\mu^{\texttt{q},\texttt{t}}(\texttt{T}_{\lambda,\texttt{t}})$ are Macdonald polynomials which are viewed as
functions of power sums variables which are $\texttt{T}_{\lambda,\texttt{t}} =
\left(T_\lambda(\texttt{t}), T_\lambda(\texttt{t}^2), T_\lambda(\texttt{t}^3),\dots \right)$ (see Remark \ref{Hall-Littlewood}).
Here the polynomials $Q^{\texttt{q},\texttt{t}}_\mu$ may be also written as symmetric functions in $d$ variables which are the
quantum contents of the diagram $\lambda$.

Let us note that the idea to weight (the $\mathbb{CP}^1$) Hurwitz numbers by symmetric functions first was worked out in
\cite{HarnadMathieu-sept-2014} where $\{h_\mu,m_\mu\}, \{e_\mu,f_\mu\}$ also $\{s_\mu\}$ and $\{p_\mu\}$ basis sets
(see \cite{Mac}) were
used. The notion of $q$-deformed Hurwitz numbers introduced in \cite{HarnadMathieu-sept-2014} in our approach
is based on  $q$-dependent specifications of the parameters $\zeta$ in the parametrization I (\ref{r-MirMorNat}), the parametrization II
 (\ref{r-new-trig}) was not considered in \cite{HarnadMathieu-sept-2014}.

(g) Remark \ref{special-q-t-limits} suggests to consider the following weighted sums of Hurwitz numbers
\[
\texttt{F}(d,\Delta,(d),\{\texttt{q}_s,\texttt{t}_s\}) :=
\sum_{\lambda\atop |\lambda|=d} \varphi_\lambda\left((d)\right)
\varphi_\lambda(\Delta)\frac{{\rm dim}\lambda}{d!}\,
\prod_{s=1}^k
\frac{s_\lambda(\bpow(\texttt{q}_s,\texttt{t}_s))}{s_\lambda(\bpow_\infty)}
\]
\be\label{F}
=\sum_{\lambda\atop{|\lambda|=d}}
\varphi_\lambda\left((d)\right)
\varphi_\lambda(\Delta)\frac{{\rm dim}\lambda}{d!}
\prod_{s=1}^k
\left(1+\sum_{\mu\neq 1^d} \varphi_\lambda(\mu) w(\Delta^{(s)}, \texttt{q}_s,\texttt{t}_s)\right)
\ee
As we see this sum describes covers with the following set of profiles on $\mathbb{RP}^2$: an arbitrary
profile $\Delta$ over $0$, the maximally ramified profile $(d)$ over another point, special weighted sums of
profiles $\Delta^{(s)},\,s=1,\dots,k$ over each of
$k$ additional branch points with the ramification weights $w(\Delta^{(s)}, \texttt{q}_s,\texttt{t}_s)$ of (\ref{w}).
(Here we skip details because
it will published in a more detailed text.
 Such sums allow to count the $d$-fold covers whose profiles $\Delta^{(s)}$ over the additional branch points
 contain given numbers of parts which are multiples of other given numbers playing the role of a chosen set of degrees of roots of unity.
 This is achieved by studying limits where the parameters $\texttt{q}_s$ and $\texttt{t}_s$ are chosen to be close to the roots of unity.)

 We will show that the numbers $\texttt{C}_\mu(d;\Delta)$, $\texttt{J}_\mu(d;\Delta)$, $\texttt{S}(d;\Delta)$ and
 $\texttt{K}_{\mu}(d;\Delta|\texttt{t})$,
 $\texttt{M}^{\texttt{q},\texttt{t}}_{\mu}(d;\Delta)$, $\texttt{F}(d,\Delta,(d),\{\texttt{q}_s,\texttt{t}_s\})$ are generated by
 special BKP tau functions considered
 in Sections \ref{Examples-tau-subsection} and \ref{BKP-tau-Hurwitz-section}. For instance the number (\ref{S}) is generated by
 (\ref{BKP-Pochhammer-texttt-a}) and the number (\ref{F}) is generated by (\ref{tau-F}) below.

\section{BKP tau functions. \label{BKP-tau-function}}

\subsection{BKP hierarchy of Kac and van de Leur.}

There are two different BKP hierarchies of integrable equations, one was introduced by the Kyoto group in
\cite{JM}, the other was
introduced by V. Kac and J. van de Leur in \cite{KvdLbispec}. We need the last one. This hierarchy includes
the celebrated
KP one as a particular reduction.  In a
certain way (see \cite{LeurO-2014}) the BKP hierarchy may be related to the three-component KP hierarchy
introduced in \cite{JM}
(earlier described in \cite{ZakharovShabat} with the help of L-A pairs of differential operators with matrix
valued coefficients).
For a detailed
description of the BKP we refer readers to the original work \cite{KvdLbispec}, and here
we write down the first non-trivial equations (Hirota equations) for the BKP tau function. These are
\bea\label{Hirota-elementary-1'}
\frac 12 \frac{\partial\tau(N,n,\bpow)}{\partial p_2} \tau(N+1,n+1,\bpow)-
\frac 12 \tau(N,n,\bpow)\frac{\partial\tau(N+1,n+1,\bpow)}{\partial p_2} \nonumber
+\frac 12 \frac{\partial^2\tau(N,n,\bpow)}{\partial^2 p_1} \tau(N+1,n+1,\bpow)\\
+\frac 12 \tau(N,n,\bpow)\frac{\partial^2\tau(N+1,n+1,\bpow)}{\partial^2 p_1}
- \frac{\partial\tau(N,n,\bpow)}{\partial p_1}\frac{\partial\tau(N+1,n+1,\bpow)}{\partial p_1}
=\tau(N+2,n+2,\bpow)\tau(N-1,n-1,\bpow)
\eea

The BKP tau functions depend on the set of higher times $t_m=\frac 1m p_m$, $m>0$ and the discrete parameter
$N$.
In \cite{OST-I}  a second discrete parameter $n$ was added, and the simplest Hirota equation relating the
BKP tau
functions for neighboring values of $n$ is
\bea\label{Hirota-elementary-2'}
\frac 12 \tau(N,n+1,\bpow)\frac{\partial^2 \tau(N+1,n+1,\bpow)}{\partial^2 p_1}-
\frac 12 \frac{\tau(N,n+1,\bpow)}{\partial^2 p_1} \tau(N+1,n+1,\bpow)=\nonumber
\\
\frac{\partial\tau(N+2,n+2,\bpow)}{\partial p_1}\tau(N-1,n,\bpow)-
 \frac{\partial \tau(N+1,n+2,\bpow)}{\partial p_1}\tau(N,n,\bpow)
\eea

 The complete set of  Hirota equations with two discrete parameters is written down
in the Appendix.

The general solution to the BKP Hirota equations may be written as
\be\label{BKP-tau-general}
\tau\left(N,n,\bpow \right)=\sum_{\lambda}\,A_\lambda(N,n) s_\lambda(\bpow)
\ee
where $A_\lambda$ satisfies the Plucker relations for an isotropic Grassmannian and (as one can show with the help of the Wick formula) may be written
in pfaffian form.

\subsection{BKP tau function of the hypergeometric type.}
We are interested in a certain subclass of the BKP tau functions (\ref{BKP-tau-general}) introduced in
\cite{OST-I} and called
BKP hypergeometric tau functions, which may be compared to a similar class of TL  and KP
tau functions found in \cite{KMMM}, \cite{OS-2000}.

Similarly to  \cite{OS-2000} we proceed as follows. Suppose that $\lambda$ is a Young diagram.
Given an arbitrary function $r$ of one variable
we construct the following product
\be
r_\lambda(x):=\prod_{(i.j)\in\lambda}\, r(x+j-i)
\ee
which is called the content product (or, sometimes, the generalized Pochhammer symbol attached to a Young
diagram $\lambda$). Examples were considered above.

\br\label{gf}
 (1) If $r=fg$, then $r_\lambda(x)=f_\lambda(x)g_\lambda(x)$.
 (2) If ${\tilde r}(x)=\left(r(x)\right)^n,\,n\in\mathbb{C}$,
then ${\tilde r}_\lambda(x)=
\left({r}_\lambda(x)\right)^n$.
\er

We consider sums over partitions of the form
\be\label{hypergBKP-sums}
\sum_{\lambda \atop \ell(\lambda)\le N} \, r_\lambda(n)\, c^{|\lambda|}\, s_\lambda(\bpow) \,=
: \tau_r^{\rm B}(N,n,\bpow)
\ee
where $s_\lambda$ are the Schur functions \cite{Mac} and $\bpow$
denotes the semi-infinite set $(p_1,p_2,\dots )$.
It was shown in \cite{OST-I} that up to a factor (\ref{hypergBKP-sums}) defines the BKP tau function:
\bp
Any given $r$ the tau function $g(n)\tau^{\rm B}_r(N,n,\bpow)$ solves the BKP Hirota equations. Here
 $g(n)$ is a function of the parameter $n$
defined by (\ref{g(n)}) in the Appendix \ref{fermionic-appendix}.
\ep
Let us mark two points: though discrete parameters enter Hirota equations, for our purposes (a) the factor $g(n)$ is
unimportant (b) the cutting $N$ should be chosen large enough and we can take $N=+\infty$.

We call such tau functions hypergeometric because both the so-called generalized
hypergeometric functions and the basic hypergeometric functions of one variable may be obtained as specifications of
(\ref{hypergBKP-sums}). For instance one can choose $p_m=x^m$. Then a rational function $r$ in (\ref{hypergBKP-sums})
yields the generalized hypergeometric function while trigonometric $r$ results in the basic one. However the key
tau function is the simplest one:

{\bf Example}. Consider $r(x)=1 $ for any $x$. Such tau function does not depend on $n$ and will be denoted
by $\tau_1(N,\bpow)$. Other hypergeometric tau functions may be obtained by action of a specially chosen vertex operator
on $\tau_1(N,\bpow)$, for example see (\ref{tau-via-vertex-t}). If we take $N=+\infty$ we can obtain
\be\label{r=1}
\tau^{\rm B}_1(\infty,\bpow)=\sum_\lambda c^{|\lambda|} s_\lambda(\bpow)=
e^{\sum_{m>0} \left(\frac {c^2}{2m}p_m^2 +c\frac {p_{2m-1}}{2m-1}\right)}
\ee

\br
Each tau function $\tau_r^{\rm B}$ may be expressed as a pfaffian, see \cite{OST-I}.
\er

\paragraph{2KP and BKP hypergeometric tau functions.}
The role of the hypergeometric functions of matrix argument in form of KP tau functions presented in \cite{OS-TMP}
was marked in \cite{Goulden-Jackson-2008} in the context of combinatorial problems.
Hypergeometric tau function of the two-component KP (2KP) may be written as
\be\label{hyperg2KP-sums}
\sum_{\lambda \atop \ell(\lambda)\le N} \, r_\lambda(n)\, c^{|\lambda|}\,s_\lambda(\bpow) \,
s_\lambda(\bbpow) \,
=
: \tau_r^{\rm 2KP}(N,n,\bpow,{\bbpow})
\ee
where $r_\lambda(n)$ is the same as in (\ref{hypergBKP-sums}).
 Here two independent sets $\bpow=(p_1,p_2,\dots)$ and $\bbpow=({\bar p}_1,{\bar p}_2,)$ and
 two discrete parameters $N$ and $n$ play the role of 2KP higher times. (We do not mark the dependence of the right hand
 side on the constant $c$ since it is trivial.)
 Then hypergeometric tau functions of
 2KP and BKP hierarchies are related:
 \be\label{2KP-to-BKP}
\left[ e^{\sum_{i>0}\frac {i}{2}\frac{\partial^2}{\partial {\bar p}_i^2}+
\sum_{i>0,\,{\rm odd}}\frac{\partial}{\partial {\bar p}_i}}\cdot
\tau_r^{\rm 2KP}(N,n,\bpow,{\bbpow})\right]_{\bbpow =0}\,=\,\tau_r^{\rm B}(N,n,\bpow)
 \ee
 which follows from
 (\ref{Laplace=sum-Schurs}) and (\ref{Schur-orthogonality}):
 \be
\left[ e^{\sum_{i>0}\frac {i}{2}\frac{\partial^2}{\partial p_i^2}+
\sum_{i>0,\,{\rm odd}}\frac{\partial}{\partial p_i}}\cdot s_\lambda(\bpow)\right]_{\bpow=0} =1,
\quad
 \ee

 \paragraph{Hypergeometric tau functions via the vertex operators.} From bosonization formulae of \cite{JM}
 in \cite{OST-I} tau functions (\ref{BKP-hyp-tau}) were presented in terms of an action of the vertex operators.
 For $r$ given by (\ref{r-new-trig}) (the same: by (\ref{choice-r-2})) the tau function (\ref{BKP-hyp-tau}) may be written as
 \be\label{tau-via-vertex-t}
 \tau_r^B(N,n,\bpow) =\,\frac{1}{g(n)}\,
 e^{\xi_0 {\hat h}_2(n) \log\texttt{t} +\sum_{m\neq 0} \texttt{p}^*_m {\hat h}(n,\texttt{t}^m)}
 \cdot \sum_{\lambda\atop \ell(\lambda)\le N}\,c^{|\lambda|}s_\lambda(\bpow)
\ee
where ${\hat h}(n,\texttt{t}^m) \, (m\in\mathbb{Z})$ are commuting operators defined as vertex operators
\be\label{hat-h-t}
{\hat h}(n,\texttt{t})\, :=\, \texttt{t}^{n}\res_z \,\frac {dz}{z}
e^{\sum_{i>0} (\texttt{t}^i-1) \frac{z^i p_i}{i}   }\,
e^{-\sum_{i>0} (\texttt{t}^{-i}-1) z^{-i} \frac{\partial }{\partial p_i} }
\ee
and where ${\hat h}_2(n)$ is determined by the generating series
${\hat h}(n,e^{\epsilon})=: 1+ \sum_{i\ge 0} \frac{\epsilon^{i+1}}{(i+1)!} {\hat h}_i(n)$.
The operators ${\hat h}_i(n)$ were written down in \cite{AMMN-2011} in the most explicit way.
From (\ref{hat-h-t}) we get
\[
{\hat h}_0(n) = n\,,\qquad {\hat h}_1(n) = n^2 + \sum_{i>0} ip_i\frac{\partial}{\partial p_i}\,,
\]
\be
{\hat h}_2(n)= n^3 + \sum_{i,j}
\left((i+j)p_ip_j\frac{\partial}{\partial p_{i+j}} + ij p_{i+j}\frac{\partial^2}{\partial p_{i}\partial p_{j}}\right)
\ee
In particular the operator ${\hat h}_2(0)$ is known as the cut-and-join operator which was introduced in \cite{GJ}.

For $r$ given by (\ref{r-MirMorNat}) (the same: by (\ref{choice-r-1})) tau function (\ref{BKP-hyp-tau}) may be written as
\[
 \tau_r^B(N,n,\bpow)=\,\frac{1}{g(n)}\,e^{\sum_{m>0} p^*_m {\hat h}_m(n) }
 \cdot \sum_{\lambda\ell(\lambda)\le N}\,c^{|\lambda|} s_\lambda(\bpow)
\]

{\bf Example.} For $N=+\infty$, $n=0$ and $r(x)=e^{\zeta_1 x}$ (that is $p^*_m=0,\,m>2$, see (\ref{xi-p*}))
 we get
\be\label{Example-vertex}
\tau^{\rm B}_r(\bpow) =\sum_{\lambda} e^{\zeta_1 \varphi_\lambda(\Gamma)}c^{|\lambda|} s_\lambda(\bpow)
=e^{\zeta_1 {\hat h}_2(0)}\cdot
e^{\sum_{m>0} \frac {c^2}{2m}p_m^2 +\frac{c}{2m-1}p_{2m-1}}
\ee

\section{Examples of the BKP hypergeometric tau functions.\label{Examples-tau-subsection}}

In case we use parameters to describe $r$, say, the parameters $\zeta$ as in (\ref{r-MirMorNat}), we shall write
$\tau^{\rm B}(N,n,\bpow|\zeta)$
instead of $\tau_r^B(N,n,\bpow)$.
Let us use Propositions \ref{Proposition-texttt-H}, \ref{Proposition-texttt-T}
 and relations (\ref{content-a}), (\ref{content-t-q}) to construct examples of BKP tau functions.
In view of (\ref{textsc-H}) each example may be considered as the generating function for certain sums
of Hurwitz numbers. More specified examples are the subject of Section \ref {BKP-tau-Hurwitz-section} where tau functions
generating sums introduced in Section \ref{Hurwitz-numbers-section}.

{\bf Example 0}. The simplest hypergeometric tau function is $\sum_{\ell(\lambda)\le N} s_\lambda(\bpow)$ is
related to $\zeta=0$.

 {\bf Example I}. First  we choose (\ref{choice-r-1}) for the content product. Using (\ref{gf})  we write
 down the following  example
\be\label{BKP-tau-zeta}
\tau^{\rm B}(N,0,\bpow|h,\zeta)\,=\,\sum_{d\ge 0}\, c^d
\sum_{\lambda \atop |\lambda|=d,\, \ell(\lambda)\le N}\,
s_\lambda(\bpow)\, \exp \sum_{m>0} \frac 1m h^m \zeta_m \Phi_m(\lambda)
\ee
If $h=1$ it may be suitable to introduce
the dependence on the variable $n$  after performing the triangle change of variables
$\zeta \to \bpow^* $
given by $V(x-1,\zeta)-V(x,\zeta)=V(x,\bpow^*)$. Then
\be\label{BKP-tau-zeta-to-p^*}
\tau^{\rm B}(N,n,\bpow|\bpow^*)\,=\,\sum_{d\ge 0} \, c^d
\sum_{\lambda\atop |\lambda|=d,\, \ell(\lambda)\le N}\,
s_\lambda(\bpow)\, \prod_{i=1}^N
e^{V(h_{i+n}(\lambda),\bpow^*)}
\ee
where $h_i(\lambda)=\lambda_i-i$.
\br

The specialization $p_m=\tr R^m=\sum_{a=1}^N x_a^m$, where  $x_i=
e^{y_i}$, allows (\ref{BKP-tau-zeta}) to be
rewritten as
  \be
 \tau^{\rm B}(N,0,\bpow|\zeta)\  =\,\frac{1}{\Delta_N({\bf x})} \sum_{h_1,\dots,h_N=1}^M\,
e^{V(h,\bpow^*)}\,\det\left(e^{y_jh_i}\right)\,\sgn \Delta_N(h)
  \ee
which is a discrete analogue of the two-matrix integral
  \be
\int \, dU \int \, dR\,\det\,R^n\,\exp\,\left(\Tr\,\left(UYU^\dag R+ \,\sum_{m\neq 0}\,\frac 1m p^*_mR^m\right) \right)
  \ee
where the first integral represents integration over unitary matrices and the second is the integral over real
symmetric ones, $dU$ and $dR$ denote the corresponding Haar measures.
$Y$ is any diagonal matrix (a source). The matrices are $N$ by $N$ ones.
This integral may be viewed as an analogue of the Kontsevich integral.

\er

{\bf Example Ia}. In (\ref{BKP-tau-zeta}) one can specify the variables $\zeta$ as
\[
 \zeta_m\,=\,\sum_{s=1}^k\,\texttt{n}_s (-\texttt{a}_s)^{-m}\,,\quad \zeta_0=-\texttt{n}_s\log \texttt{a}_s
\]
where $\texttt{a}_s\in\mathbb{C}$.
If we restore the dependence of tau function on $n$ we obtain
\be\label{BKP-Pochhammer-texttt-a}
\tau^{\rm B}(N,n,\bpow|h,\{ \texttt{a}_s,\texttt{n}_s \})\,=\,\sum_{d\ge 0}\, c^d
\,\sum_{\lambda\atop |\lambda|=d,\, \ell(\lambda)\le N}\,s_\lambda(\bpow) \prod_{s=1}^k \,\prod_{(i.j)\in\lambda}\,
\left(1+h\frac{n+j-i}{a_s} \right)^{-\texttt{n}_s}
\ee
where $\texttt{a}$ and $\texttt{n}$ are respectively the collections of complex parameters
$\texttt{a}_1,\dots,\texttt{a}_k$ and $\texttt{n}_1,\dots,\texttt{n}_k$. For $\texttt{n}_s=\pm$ we obtain the
pfaffian version of the hypergeometric function of matrix argument \cite{OS-TMP}.

{\bf Example Ib}. Let us take all $\texttt{n}_s$ equal to $\texttt{n}(\alpha)=\frac{1}{\alpha}$ in the previous example. We obtain
\be\label{BKP-Pochhammer-Jack}
\tau^{\rm B}(N,n,\bpow|h,\texttt{a},\texttt{n}(\alpha))\,=\,\sum_{d\ge 0}\, c^d
\,\sum_{\lambda\atop |\lambda|=d,\, \ell(\lambda)\le N}\,s_\lambda(\bpow) \sum_{\mu} h^{|\mu|}
P_\mu^{(\alpha)}(-\texttt{a}(n))Q_\mu^{(\alpha)}\left(\Phi(\lambda)\right)
\ee
where $P_\mu^{\alpha}$ and $Q_\mu^{\alpha}$ is the pair of dual Jack polynomials written in  the notation of Ch IV \cite{Mac}.
Here the first Jack polynomial, $P_\mu^{\alpha}$, is a
symmetric function of the variables
$-\texttt{a}(n) = (-a_1-n,\dots,-a_k-n)$ while the second Jack polynomial $Q_\mu^{\alpha}\left(\Phi(\lambda)\right)$ should be viewed as
 quasihomogeneous polynomial in power sum variables $\Phi=(\Phi_1(\lambda),\Phi_2(\lambda),\dots)$.

{\bf Example II}. Next we use (\ref{choice-r-2}) and (\ref{gf}) getting
\be\label{BKP-tau-Okounkov-AMMN-q-i}
\tau^{\rm B}(N,n,\bpow|\{{\bf \texttt{p}}^{*(s)},\texttt{t}_s\})=
\sum_{d\ge 0}\, c^d\sum_{\lambda\atop |\lambda|=d,\, \ell(\lambda)\le N}\,s_\lambda(\bpow)\prod_{s=1}^k
e^{\xi_0 (\varphi_\lambda(\Gamma)+nd)\log \texttt{t}_{s}+\sum_{m\neq 0} \xi_m^{(s)} \texttt{t}_s^{mn}
T_\lambda(\texttt{t}_s^m) }
\ee
\be
\label{BKP-tau-Okounkov-AMMN-q-i-t^*}
=\sum_{d\ge 0}\, c^d\sum_{\lambda\atop |\lambda|=d,\, \ell(\lambda)\le N}\,s_\lambda(\bpow)\prod_{s=1}^k
e^{\xi_0 (\varphi_\lambda(\Gamma)+nd)\log\texttt{t}_s - \sum_{m\neq 0} \frac 1m (1-\texttt{t}_s^{m}) \texttt{p}^{*(s)}_m
\texttt{t}_s^{mx-m} T_\lambda(\texttt{t}_s^m)}
\ee
 The variables ${\bf\texttt{p}}^{*(s)}$ are
related to the variables $\xi^{(s)}$ by $\texttt{p}^*_m=\xi_m\frac{\texttt{t}^m}{\texttt{t}^m-1}$.

For $k=1$ (here we will re-denote ${\bf\texttt{p}}^{*(1)}\to -{\bf \texttt{p}}^*$) and $\texttt{p}^*_m=0,\,m<0$ we have
\bea
\label{BKP-via-Hall-Littlewood}
\tau^{\rm B}(N,n,\bpow|{\bf \texttt{p}}^*,{\bf t})=\sum_{d\ge 0}\, c^d
\,\sum_{\lambda\atop |\lambda|=d,\, \ell(\lambda)\le N}\,s_\lambda(\bpow)\,\prod_{s=1}^k
{\texttt{t}}^{\xi_0 \varphi_\lambda(\Gamma)+d\xi_0}\sum_{\mu}\,
P^{\texttt{0},\texttt{t}}_\mu({\bf \texttt{p}}^*) Q^{\texttt{0},\texttt{t}}_\mu(T_{\lambda,\texttt{t}})
\eea
where $P^{\texttt{0},\texttt{t}}_\lambda$ and $Q^{\texttt{0},\texttt{t}}_\lambda$ are the
Macdonald polynomials specified by $\texttt{q}=0$  (Hall-Littlewood polynomials)
 may be written either as the quasihomogeneous polynomial of the
power sum variables $T_{\lambda,\texttt{t}}=(T_\lambda(\texttt{t}),T_\lambda(\texttt{t}^2),\dots)$, or as the symmetric polynomial in quantum
contents, see Remark \ref{Hall-Littlewood}.

\br\label{circular-beta=1}
Given $s$ let us specify $\bpow=\bpow(\texttt{q},\texttt{t})$ according to (\ref{p(t,q)}). Then
the series (\ref{BKP-tau-Okounkov-AMMN-q-i}) solves the BKP Hirota equations with respect to the variables
${\bf\texttt{p}}^{*}$. In case $|\texttt{t}|=1$ and is not a root of 1,
$\tau^{\rm B}$ of (\ref{BKP-tau-Okounkov-AMMN-q-i})
is basically a discrete version of the circular $\beta=1$ ensemble
\[
 \frac{1}{N!} \sum_{h_1,\dots,h_N}\,\prod_{i<j}|\texttt{t}^{h_i}-\texttt{t}^{h_j}| \,
 \prod_{i=1}^N e^{V({\bf\texttt{p}}^{*},\texttt{t}^{h_i})} \mu(h_1;\texttt{q},\texttt{t})
\]
with a certain weight function $\mu$ independent of ${\bf \texttt{p}}^*$,
see \cite{OST-I}. This may be compared to Remark \ref{remark-kontsevich} and to the discrete version
of the orthogonal ensemble (\ref{discrete-orthogonal}).
\er

Consider three specifications of the variables $\xi$ in (\ref{BKP-tau-Okounkov-AMMN-q-i}).

{\bf Example IIa}. First, we put each $\xi^{(s)}_m=0$, $s=1,\dots,k$. Then the content product depends only
on the parameter $\xi_0$. We obtain the BKP analogue of the Okounkov's TL tau function presented in \cite{Okounkov-2000}:
\be\label{BKP-Okounkov}
\tau^{\rm B}(N,n,\bpow |\xi_0)\,=\,\sum_{d\ge 0} \,c^d
\sum_{\lambda\atop |\lambda|=d,\, \ell(\lambda)\le N}\,s_\lambda(\bpow)\,
\prod_{(i.j)\in\lambda} e^{(n+j-i) \xi_0 }
\ee

{\bf Example IIb}. Now, take $\xi_0=0$ and
\be
\label{xi-for-trig-Pochhamer}
 \xi_m^{(s)}  \,=\,\frac{\texttt{t}^m-1}{\texttt{t}^m}\texttt{p}_m^{*(s)}=\,\, \texttt{n}_s \texttt{q}_s^{m} \,,\quad m>0
\ee
We obtain
\be\label{BKP-tau-trig-Pochhammer-n-i}
\tau^{\rm B}(N,n,\bpow | \{\texttt{t}_s,\texttt{q}_s,\texttt{n}_s \} )\,=\,\sum_{d\ge 0}\, c^d
\sum_{\lambda\atop |\lambda|=d,\, \ell(\lambda)\le N}\,s_\lambda(\bpow)\,
\prod_{s=1}^k\,\prod_{(i.j)\in\lambda}\left( 1-\texttt{q}_s\texttt{t}_s^{n+j-i} \right)^{-\texttt{n}_s}
\ee
where $\texttt{t},\texttt{q},\texttt{n}$ are sets of complex numbers
$\texttt{t}_s,\texttt{q}_s,\texttt{n}_s$, $s=1,\dots,k$.

In case $\texttt{n}_s=\pm 1$, $s=1,\dots,k$ the tau function (\ref{BKP-tau-trig-Pochhammer-n-i}) is the pfaffian
version of Milne's hypergeometric function \cite{Milne}.

{\bf Example IIc}. Next, take $\xi_0=0$ and
\[
 \xi_{\pm m}^{(s)}
 =\,\frac{\texttt{t}^{\pm m}-1}{\texttt{t}^{\pm m}}\texttt{p}_{\pm m}^{*(s)}=
 (-1)^{m}\texttt{n}_s\frac {\texttt{q}_s^{\frac m2} \texttt{t}_s^{\pm\texttt{a}_sm}}{1-\texttt{q}_s^m}\,\,
 \quad s=1,\dots,k,\quad m>0
\]
and put $\texttt{q}_s=e^{2\pi i \tau_s},\, \texttt{t}_s=e^{2c_s\pi i}$.
Then (\ref{BKP-tau-Okounkov-AMMN-q-i}) takes the form
\be\label{BKP-tau-elliptic-Poch}
\tau^{\rm B}(N,n,\bpow |\{\texttt{c}_s,\tau_s,\texttt{a}_s,\texttt{n}_s\})\,=\,\sum_{d\ge 0} \, c^d
\sum_{\lambda\atop |\lambda|=d,\, \ell(\lambda)\le N}\,s_\lambda(\bpow)\,
\prod_{s=1}^k\,\left(\theta_\lambda(\texttt{c}_s(n+\texttt{a}_s),\tau_s) \right)^{-\texttt{n}_s}
\ee
where  $\{\texttt{c},\tau,\texttt{a},\texttt{n}\}$ are sets of complex numbers
$\{\texttt{c}_s,\tau_s,\texttt{a}_s,\texttt{n}_s$, $s=1,\dots,k\}$, and where
\[
 \theta_\lambda(c_s(n+\texttt{a}_s),\tau_s) :=
 \prod_{(i.j)\in\lambda}\,\theta(c_s(n+\texttt{a}_s+j-i),\tau_s)
\]
is the elliptic version of the Pochhammer symbol, $\theta$ is the Jacobi theta function
\[
 \theta(c_s x,\tau_s):=\sum_{k\in\mathbb{Z}}\exp (\pi i k^2 \tau_s + 2c_s\pi i k x)=
 (\texttt{q}_s;\texttt{q}_s)_\infty
\prod_{k=1}^\infty \left(1 + \texttt{q}_s^{k-{\frac 12}} \texttt{t}_s^{ x}\right)\left(1 +
\texttt{q}_s^{k-{\frac 12}} \texttt{t}_s^{- x}\right)
\]
where $(\texttt{q}_s;\texttt{q}_s)_\infty$ is the Dedekind function. For this example we chose
$c=(\texttt{q}_s;\texttt{q}_s)_\infty$
in (\ref{BKP-tau-Okounkov-AMMN-q-i}).
For $\texttt{n}_s=\pm 1$ we obtain the pfaffian version
of an elliptic hypergeometric function considered in \cite{OS-2000}.

{\bf Example IId}. In (\ref{BKP-via-Hall-Littlewood}) we choose $k=1$, $\texttt{n}=1$. Let us take take
\[
 \xi_{ m}\,=\,\frac {1-\texttt{t}^m}{1-\texttt{q}^m}\sum_{i=1}^k y_i^m\,,
 \quad m>0
\]
all other variables vanish. This may be viewed as a limiting case
 of the previous Example Ib where we send $k\rightarrow\infty$.  Then
\[
r(x)= \prod_{m>0}\prod_{i=1}^k \,\frac{1-y_i q^m \texttt{t}^{x+1}}{1-y_i q^m \texttt{t}^{x}}
\]
The content product is equal to
\be
 \prod_{(i.j)\in\lambda}\,\prod_{m>0}\,\prod_{i=s}^k \,\frac{1-y_i q^m \texttt{t}^{x+1+j-i}}{1-y_s q^m \texttt{t}^{x+j-i}}=
e^{
\sum_{m > 0} \frac 1m \frac{1-\texttt{t}^{m}}{1-\texttt{q}^{m}} \texttt{t}^{mx} T_\lambda(\texttt{t}^m)\sum_{i=1}^k y_i^m}=
\sum_{\mu}\, \texttt{t}^{x|\mu|} P^{\texttt{q},\texttt{t}}_\mu(Y) Q^{\texttt{q},\texttt{t}}_\mu(T_{\lambda,\texttt{t}})
\ee
where the Macdonald function $P^{\texttt{q},\texttt{t}}_\mu$ is the symmetric polynomial in
$Y=(y_1,\dots,y_k)$ and the Macdonald function $Q^{\texttt{q},\texttt{t}}_\mu$ may be written either as the
quasihomogeneous polynomial of the
power sum variables $T_{\lambda,\texttt{t}}=(T_\lambda(\texttt{t}),T_\lambda(\texttt{t}^2),\dots)$, or as the symmetric
polynomial in quantum
contents,
see Remark \ref{Hall-Littlewood}.
The tau function (\ref{BKP-via-Hall-Littlewood}) takes the form
\bea\label{Macdonald}
\tau^{\rm B}(N,n,\bpow |\texttt{q},\texttt{t},\xi_0,Y) \,&=&\, \sum_{d\ge 0} \,c^d
\sum_{\lambda\atop |\lambda|=d,\, \ell(\lambda)\le N} s_\lambda(\bpow)
\,e^{\xi_0 \varphi_\lambda(\Gamma)}
\sum_{\mu}\, \texttt{t}^{n|\mu|} P^{\texttt{q},\texttt{t}}_\mu(Y)
Q^{\texttt{q},\texttt{t}}_\mu(T_{\lambda,\texttt{t}})\qquad\qquad
\\
&=&\,
\sum_{d\ge 0}\,c^d\sum_{\lambda\atop |\lambda|=d,\, \ell(\lambda)\le N}
\prod_{j=1}^N e^{\xi_0 (\lambda_j -j+n)^2}\,s_\lambda(\bpow)\,\prod_{j=1}^{N}\,\prod_{i=1}^k\,
\prod_{m>0} e^{\frac{y_i^m}{1-q^m}\texttt{t}^{m(\lambda_j-j+n-1)}}
  \label{s-s-s}
\eea
where $P^{\texttt{q},\texttt{t}}_\mu$ and $Q^{\texttt{q},\texttt{t}}_\mu$ are Macdonald polynomials, see
Remark \ref{Hall-Littlewood}. The last equality follows from (\ref{p^*-t^h}).

{\bf Example III}. Let us choose
\[
 r(x)=(a+x)\prod_{s=1}^k \frac{1-\texttt{q}_s\texttt{t}_s^x}{1-e^{\epsilon_s}\texttt{t}_s}(a_s+x)
\]
where we used both parameterizations (see Remark \ref{gf}).
We obtain the following tau function
\be\label{tau-F}
 \tau^{\rm B}(N,n,\bpow |a,\{a_s,\texttt{t}_s,\epsilon_s  \} ) =\sum_{d\ge 0}c^d\sum_{\lambda\atop |\lambda|=d } s_\lambda(\bpow)
 \frac{s_\lambda(\bpow(a))}{s_\lambda(\bpow_\infty)}\prod_{s}
 \frac{s_\lambda(\bpow(\texttt{q}_s,\texttt{t}_s))}{s_\lambda(\bpow(e^{\epsilon_s},\texttt{t}_s))}
 \frac{s_\lambda(\bpow(a_s))}{s_\lambda(\bpow_\infty)}
\ee
In particular these tau functions generate sums $\texttt{F}(d,\Delta,(d),\{\texttt{q}_s,\texttt{t}_s\})$ of (\ref{F}).

\br
Formula (\ref{BKP-Pochhammer-texttt-a}) may be obtained as a limiting case of (\ref{BKP-tau-trig-Pochhammer-n-i})
if we take $\texttt{q}_s=\texttt{t}_s^{\texttt{a}_s}$ and send $\texttt{t}\to 1$ taking into account
that for the hypergeometric tau functions  (\ref{hypergBKP-sums}) there is the obvious transformation
$r_\lambda \to a^{-|\lambda|}r_\lambda,\,p_m \to ap_m,\,m>0$,  which leaves them unchanged.

In this limiting case polynomials $P^{\texttt{q},\texttt{t}}$ and $Q^{\texttt{q},\texttt{t}}$
goes to Jack polynomials \cite{Mac}, compare to (\ref{BKP-Pochhammer-Jack}).
\er

\br \label{E=1 to E=0} Similarly to (\ref{2KP-to-BKP})
 one may prove the relation
\be\label{E=1--E=0}
 e^{\frac 12 \sum_{m>0} m \frac{\partial^2}{\partial p_m^2}+\sum_{m>0,\,{\rm odd}}
 \frac{\partial}{\partial p_m}}   \cdot\tau^{\rm B}_r(N,n,\bpow)|_{\bpow=0}=
 \sum_{d\ge 0}\, c^d \sum_{\lambda\atop |\lambda|=d,\,  \ell(\lambda)\le N}\,\prod_{(i.j)\in\lambda}r(n+j-i)
\ee
where the right hand side generates weighted Hurwitz numbers for the torus and the Klein bottle.
\er

\section{BKP tau functions generating Hurwitz numbers \label{BKP-tau-Hurwitz-section}}

\subsection{Hurwitz numbers themselves}

As we shall see the hypergeometric tau functions generate weighted sums of Hurwitz numbers.
However there exist special cases when one gets Hurwitz numbers themselves, this is based on Remark \ref{special-cases}.

We will make difference between the parameterizations I and II.

First, let us write down the simplest case of a single branch point related to all $r=1$ and $N=\infty$.
This case is generated by $\tau_1^{\rm B}$ where it is reasonable to produce the change $p_m\to h^{-1}c^m p_m$.
We get
\be\label{single-branch-point}
e^{\frac {1}{h^2}\sum_{m>0} \frac {1}{2m}p_m^2 c^{2m} +\frac 1h\sum_{m {\rm odd}} \frac 1m p_m c^m}=
\sum_{d>0} c^d \sum_{\textsc{e}' \le d} h^{-\textsc{e}'}\sum_{\Delta\atop |\Delta|=d,\,\ell(\Delta)=\textsc{e}'}\bpow_\Delta
H^{1,1}(d;\Delta)
\ee
where $H^{1,1}(d;\Delta)$ is the Hurwitz number describing $d$-fold covering of $\mathbb{RP}^2$ with a single
branch point of type $\Delta=(d_1,\dots,d_l),\,|\Delta|=d$ by a (not necessarily connected) Klein surface of
Euler characteristic $\textsc{e}'=\ell(\Delta)$. For instance, for $d=3$, $\textsc{e}'=1$ we get
$H^{1,1}(\Delta)=\frac 13\delta_{\Delta,(3)}$.
In particular for unbranched covering we get formula (\ref{unbranched}).

Next let us notice that the exponent of the left hand side may be rewritten as the generating series of the
connected Hurwitz numbers
\[
 \frac{1}{h^2}\sum_{d=2m}c^{2m} p_m^2 H_{\rm con}^{1,1}\left(d;(m,m)\right) +
 \frac{1}{h} \sum_{d=2m-1} c^{2m-1}p_{2m-1} H_{\rm con}^{1,1}\left(d;(2m-1)\right)
\]
where $H_{{\rm con}}^{1,1}$ describes $d$-fold covering either by the Riemann
sphere ($d=2m$) or by the projective plane ($d=2m-1$). These are the only ways to cover $\mathbb{RP}^2$
by a connected surface for the case of the single branch point.
The geometrical meaning of the exponent in (\ref{single-branch-point}) may be explained as follows. The projective plain may be viewed as the unit disk with the identification
of the opposite points $z$ and $-z$ on the boarder: $|z|=1$. In case we cover the Riemann sphere by the Riemann sphere $z\to z^m$ we get
two critical points with the same profiles. However we cover $\mathbb{RP}^2$ by the Riemann sphere, then we have the composition of the
mapping $z\to z^{m}$ on the
Riemann sphere and the factorization by antipodal involution $z\to - \frac{1}{\bar z}$. Thus we have the ramification profile $(m,m)$
at the single critical point $0$ of $\mathbb{RP}^2$.
The automorphism group is the dihedral group of the order $2m$ which consists of rotations on $\frac{2\pi }{m}$ and antipodal involution
$z\to -\frac{1}{\bar z}$.
Thus we get that $H_{\rm con}^{1,1}\left(d;(m,m)\right)=\frac{1}{2m}$ which is the factor in the first sum in the exponent in
(\ref{single-branch-point}). Now let us cover $\mathbb{RP}^2$ by $\mathbb{RP}^2$ via $z\to z^d$. For even $d$ we have the critical point
$0$, in addition each point of the unit
circle $|z|=1$ is critical (a folding), while from the beginning we restrict our consideration only on isolated critical points.
For odd $d=2m-1$ there is
the single critical point $0$, the automorphism group consists of rotations on the angle $\frac{2\pi}{2m-1}$. Thus in this case
$H^{1,1}\left(d;(2m-1)\right)=\frac{1}{2m-1}$ which is the factor in the second sum in the exponent in (\ref{single-branch-point}).

Next, let us consider BKP hypergeometric function in the parametrization I where we put
$\zeta_1=\beta +\sum_{i=1}^L a_i^{-1} $ and $\zeta_k=\sum_{i=1}^L a_i^{-k} $
\be\label{tau-Hurwitz-themselves}
H^{1,b+c+1}(d; \underbrace{\Gamma,\dots,\Gamma}_b,\underbrace{(d),\dots,(d)}_m,\Delta )= c^{-d}h^{\textsc{e}'}
\left[ \tau(N=+\infty,0,\bpow|\zeta) \right]_{b,m}
\ee
where the brackets $[*]_{b,m}$ means the picking up the factor of the term $\bpow_\Delta \prod_{i=1} a_i^{-1}$
which counts $d$-fold covers of $\mathbb{RP}^2$ with the following ramification type: there are $b$ simple branch points,
$m$ branch points of type $(d)$ and one branch point of type $\Delta=(d_1,\dots,d_l)$. Each cover is a
connected Klein surface in case $m>0$ and
not necessarily connected one in case $m=0$ )

of Euler characteristic
$\textsc{e}'=d-b -m(d-1)$. See Proposition \ref{d-cycle-proposition} on the properties of Hurwitz numbers regarding
branch points with the profiles $(d)$ (maximally ramified branch points).

\subsection{BKP tau function as generating function for the weighted sums of Hurwitz numbers.
\label{BKP tau function as the generating function}}

In this subsection the power of $\frac 1h$ counts the Euler characteristic of the covering surface denoted
by $\textsc{e}'$.

For this purpose we change
\be\label{rescale-p}
p_m=h^{-1}{\tilde p}_m
\ee

First of all we present the simplest weighted sum of Hurwitz numbers which is just the sum of Hurwitz numbers
related to a single branch point with a fixed Euler characteristic $\textsc{e}'$

\be
\sum_{d\ge 0} \,c^d   \sum_{\ell} \, h^{d-l} \sum_{\Delta^{(i)}\atop \ell(\Delta^{(i)})=l} H^{1,1}(d;\Delta^{(i)} ) =
\frac{1}{(1-c^2)^{h^{-2}}}\left(\frac{1+c}{1-c}\right)^{h^{-1}}
\ee
where each $\Delta^{(i)}$ has the same weight $d$ and length $\ell$.

From previous sections we found

\bp \label{Theorem-BKP-zeta}
The tau function (\ref{BKP-tau-zeta}) generates the  numbers $\texttt{C}_\mu(\Delta)$ (\ref{texttt-H-mu})
through
\be\label{tau-BKP-zeta-Hurwitz}
\tau^{\rm B}(N,0,\bpow|h,\zeta)\,=\,\sum_{d\ge 0}\, c^d \sum_{\mu,\Delta\atop |\Delta|=d}\, h^{|\mu|-\ell(\Delta)}
\frac{1}{z_\mu}\,\texttt{C}_\mu(\Delta) \,\zeta_\mu \,\bpow_{\Delta}
\ee
where $z_\mu$ is defined by (\ref{C-Delta,z-Delta}).
For $d=|\Delta|\le N$ the numbers $\texttt{C}_\mu(\Delta)$ are weighted Hurwitz numbers.
\ep

\begin{Corollary}
In particular, let us put $\zeta_m=0$ if $m>1$. Then (\ref{tau-BKP-zeta-Hurwitz}) reads
\be\label{Okounkov-series-RP-2}
\sum_{d\ge 0}\, c^d\sum_{\lambda\atop{|\lambda|=d,\,\ell(\lambda)\le N}}  e^{h\zeta_1 \varphi_\lambda(\Gamma)} s_\lambda(\bpow)  =
\sum_{d,b\ge 0} \,c^d  \sum_{\Delta\atop |\Delta|=d} h^{b-\ell(\Delta)}
{\tilde \bpow}_\Delta \frac{\zeta_1^{b}}{b!}   H(d;
 \underbrace{\Gamma,\dots,\Gamma }_{b} ,\Delta)
 \ee
 which is the $\mathbb{RP}^2$ analogue of the Okounkov generating function \cite{Okounkov-2000}.
\end{Corollary}
The representation of this series in the form of a matrix integral is written down below,
see (\ref{Okounkov-tau-normal-matrices-BKP}).

Weighted sums of Hurwitz numbers generated by the BKP tau functions (\ref{BKP-tau-trig-Pochhammer-n-i}) and
(\ref{BKP-Pochhammer-texttt-a}) were written down in our previous work \cite{NO-2014}. The simplest
example resulting from (\ref{BKP-Pochhammer-texttt-a}) is similar to one considered in \cite{HO-2014},
and may be presented as follows.
The tau function (\ref{BKP-Pochhammer-texttt-a}),with
$\texttt{n}_s=1$ for $s=1,\dots,k$, generates sums $S$ defined by (\ref{S}):
\bp  For the discussion compare to Example 2.22 in \cite{KazarianLando}
\[
\tau^{\rm B}(N,n,\bpow|h,\{a_s \})=\,\,
\sum_{d\ge 0}\,c^d\sum_{\lambda\atop |\lambda|=d,\, \ell(\lambda)\le N}\,s_\lambda(\bpow) \prod_{s=1}^k \,\prod_{(i.j)\in\lambda}\,
\left(a_s h^{-1}+n+j-i \right)
\]
\be
=\,\sum_{d\ge 0}\,c^d\sum_{\Delta\atop |\Delta|=d}\sum_{\mu}
\frac{1}{d!}\,(a_sh^{-1}+n)^{d-\mu_s} \bpow_{\Delta}\,S_{\mu}(d,\Delta)
\ee
\ep

\bp \label{Theorem-BKP-xi}
The tau function (\ref{BKP-tau-Okounkov-AMMN-q-i}) generates the  numbers $\texttt{K}_{\mu^{(s)}}(\Delta|\texttt{t}_s)$
(\ref{texttt-T-mu}) through
\be
\tau^{\rm B}(N,n,\bpow|\xi,\{ \texttt{t}_s\})\,=\,\sum_{d\ge 0} \, c^d
\sum_{\mu,\Delta\atop |\Delta|=d}\,\prod_{s=1}^k \,\frac{1}{d!z_\mu}\,
\bpow_{\Delta}\,\xi_{\mu^{(s)}} \, \texttt{K}_{\mu^{(s)}}(\Delta|\texttt{t}_s)
\ee
where $z_\mu$ is defined by (\ref{C-Delta,z-Delta}).
For $d=|\Delta|\le N$ the numbers $\texttt{K}_{\mu^{(s)}}(\Delta|\texttt{t}_s)$ are weighted Hurwitz numbers.
\ep

\bp
The tau function (\ref{Macdonald}) generates the Hurwitz numbers $\texttt{M}^{\texttt{q},\texttt{t}}_\mu$
weighted by Macdonald polynomials (see (\ref{texttt-M-mu})):
\be\label{prop3}
\tau^{\rm B}(N,n,\bpow |\texttt{q},\texttt{t},0,Y) \,=\, \sum_{d\ge 0}\,c^d\sum_{\Delta\atop |\Delta|=d} \,\frac{1}{d!}\,\bpow_\Delta
\sum_{\mu}\, \texttt{t}^{n|\mu|} P^{\texttt{q},\texttt{t}}_\mu(Y) \texttt{M}^{\texttt{q},\texttt{t}}_\mu(d;\Delta)
\ee
\ep

\bp
The numbers $\texttt{F}(d,\Delta,(d),\{\texttt{q}_s,\texttt{t}_s \})$  given by (\ref{F}) may be obtained as the following term
in the tau function (\ref{F}):
\be\label{tau-F'}
 \tau^{\rm B}(N,n,\bpow |a,\{a_s,\texttt{t}_s,\epsilon_s   \} ) =\sum_{d\ge 0}\, c^d\sum_\lambda s_\lambda(\bpow) \frac{s_\lambda(\bpow(a))}{s_\lambda(\bpow_\infty)}\prod_{s=1}^k
 \frac{s_\lambda(\bpow(\texttt{q}_s,\texttt{t}_s))}{s_\lambda(\bpow(e^{\epsilon_s},\texttt{t}_s))}
 \frac{s_\lambda(\bpow(a_s))}{s_\lambda(\bpow_\infty)}
\ee
\[
=\sum_{d\ge 0}\,c^d \sum_{\Delta\atop|\Delta|=d} \bpow_\Delta
\left( \texttt{F}(d,\Delta,(d),\{\texttt{q}_s,\texttt{t}_s\})\,a\prod_{s=1}^k \frac{a_s}{\epsilon_s} +\dots \right)
 \]
 where  dots means terms of different order in any of $\epsilon_s$, $a_s$ ($s=1,\dots,k$) and $a$.

\ep

\section{Matrix integrals as generating functions of Hurwitz numbers
\label{Matrix-integrals}}

In case the base surface is $\mathbb{CP}^1$ the set of examples of matrix integrals generating Hurwitz numbers were studied in
works \cite{Chekhov-2014},\cite{MelloKochRamgoolam},\cite{AMMN-2014},\cite{ChekhovAmbjorn},\cite{KZ},\cite{ZL},\cite{Zog}.
One can show that the perturbation series in coupling constants of these integrals (Feynman graphs) may be related to TL
(KP and two-component KP)
hypergeometric tau functions. It actually means that these series generate Hurwitz numbers with at most two arbitrary profiles
while other ones are subjects of certain conditions since the origin of additional profiles is the content product factors in
hypergeometric tau functions (\ref{hyperg2KP-sums}).

Here, very briefly, we will write down few generating series for the $\mathbb{RP}^2$ Hurwitz numbers.
These series may be not tau functions themselves but may be presented as integrals of tau functions of matrix argument.
(The matrix argument, which we denote by a capital letter, say $X$, means that the power sum variables $\bpow$ are specified
as $p_i=\tr X^i,\,i>0$. Then instead of
$s_\lambda(\bpow)$, $\tau(\bpow)$ we write $s_\lambda(X)$ and $\tau(X)$). If a matrix integral in examples below is a BKP tau
function then it generates Hurwitz numbers with a single arbitrary profile and all other are subjects of restrictions
identical to those in $\mathbb{CP}^1$ case mentioned above.
In all examples $V$ is given by (\ref{V}). We also recall that the limiting values of $\bpow(\texttt{q},\texttt{t})$
given by (\ref{p(t,q)}) may be $\bpow(a)=(a,a,\dots)$ and $\bpow_\infty=(1,0,0,\dots)$. We also recall that numbers
$H^{\textsc{e},\textsc{f}}(d;\dots)$ are Hurwitz numbers only in case $d\le N$, $N$ is the size of matrices.

For more details of the $\mathbb{RP}^2$ case see \cite{NO-2014}. New development in \cite{NO-2014} with respect to
the consideration in \cite{O-2002} is the usage of products of matrices. 
Here we shall consider a few examples. 
All examples include the simplest BKP tau function, of matrix argument $X$,
\cite{OST-I} defined by (compare to (\ref{sum-Schurs}))
\be\label{vac-tau-BKP'}
 \tau_1^{\rm B}(X)\,:=\,\sum_\lambda \,s_\lambda(X)\,=
 e^{\frac 12 \sum_{m>0} \frac 1m\left(\tr X^m\right)^2 + \sum_{m>0,{\rm odd}}\frac{1}{m}\tr X^m}
= \frac{\det^{\frac 12}\frac{1+X}{1-X} }{\det^{\frac 12}\left( I_N \otimes I_N - X\otimes X\right)}
\ee
as the part of the integration measure. Other integrands are the simplest KP tau functions
$\tau_1^{\rm KP}(X,\bpow):=e^{\tr V(X,\bpow)}$ where $V$ is defined by (\ref{V}) where the parameters
$\bpow$ may be called coupling constants. The perturbation series in coupling constants are expressed
as sums of products of the Schur functions over partitions and are similar to the series we considered in
the previous sections.

{\bf Example 1. The $\mathbb{RP}^2$ Okounkov Hurwitz series as a model of normal matrices.}
From the equality
\[
\left({2\pi}{\zeta_1^{-1}} \right)^{\frac 12} e^{\frac{(n\zeta_0)^2}{2\zeta_1}} e^{\zeta_0 nc+ \frac12 \zeta_1 c^2}\,
=\,
 \int_{\mathbb{R}} e^{x_i n\zeta_0 +(cx_i- \frac12 x^2_i)\zeta_1} dx_i ,
\]
 in a similar way as was done in \cite{OShiota-2004} using $\varphi_\lambda(\Gamma)=\sum_{(i.j)\in\lambda}(j-i)$,
 one can derive
\[
 e^{n|\lambda|\zeta_0}e^{\zeta_1 \varphi_\lambda(\Gamma)}\delta_{\lambda,\mu}\,=\,\textsc{k} \,
 \int  s_\lambda(M) s_\mu(M^\dag) \det \left(MM^\dag\right)^{n\zeta_0}
 e^{-\frac12 \zeta_1\tr \left( \log \left( MM^\dag\right)\right)^2} dM
\]
where $\textsc{k}$ is unimportant multiplier,  $M$ is a normal matrix with eigenvalues $z_1,\dots,z_N$ and $\log |z_i|=x_i$,
and
$dM=\,d_*U\,\prod_{i<j}|z_i-z_j|^2\prod_{i=1}^N d^2 z_i$. Then the $\mathbb{RP}^2$ analogue of the Okounkov series
(\ref{Okounkov-series-RP-2}) may be written
\be\label{Okounkov-tau-normal-matrices-BKP}
\sum_{\lambda\atop \ell(\lambda)\le N}e^{n|\lambda|\zeta_0 +
\zeta_1 \varphi_\lambda(\Gamma)}
s_\lambda(\bpow)=\textsc{k}
 \int  e^{V(M,\bpow)}
 e^{\zeta_0 n\tr \log \left(MM^\dag\right)-\frac12 \zeta_1\left( \tr\log \left( MM^\dag\right)\right)^2}
 \tau^{\rm B}_1(M^\dag) dM
\ee

A similar representation of the Okounkov $\mathbb{CP}^1$ series  was earlier presented in
\cite{AlexandrovZabrodin-Okounkov}.

Below we use the following notations
 \begin{itemize}
  \item $  d_*U $ is the normalized Haar measure on $\mathbb{\mathbb{U}}(N)$: $\int_{\mathbb{U}(N)}d_*U =1$

  \item $Z$ is a complex matrix
    $$
d\Omega(Z,Z^\dag)  =\,\pi^{-n^2}\,e^{-\tr \left(ZZ^\dag\right)}\,
\prod_{i,j=1}^N \,d \Re Z_{ij}d \Im Z_{ij}
  $$

  \item Let $M$ be a Hermitian matrix the measure is defined
   $$
   dM= \, \prod_{i\le j}
d\Re M_{ij} \prod_{i<j} d\Im M
  $$

 \end{itemize}

It is known \cite{Mac}
\be\label{s-s-N_lambda-1}
\int s_\lambda(Z)s_\mu(Z^\dag)\,d\Omega(Z,Z^\dag) = (N)_\lambda\delta_{\lambda,\mu}
\ee
where $(N)_\lambda:=\prod_{(i.j)\in\lambda}(N+j-i)$ is the Pochhammer symbol
related to $\lambda$. A similar relation
was used in \cite{O-Acta},\cite{HO-2MM},\cite{O-2002},\cite{AMMN-2014},\cite{OShiota-2004}, for models of Hermitian, complex
and normal matrices.

By $I_N$ we shall denote the $N\times N$ unit matrix.
We  recall that
$$ s_\lambda(I_N)=(N)_\lambda s_\lambda(\bpow_\infty)\,,
\qquad s_\lambda(\bpow_\infty) = \frac{{\rm dim}\lambda}{d!},\quad d=|\lambda|$$.

{\bf Example 2. Three branch points.}
The generating function for $\mathbb{RP}^2$ Hurwitz numbers with three ramification points,having two
arbitrary profiles at $0$ and at $\infty$ with fixed length in the third point:
\be\label{3-points-integral}
 \sum_{\lambda}\frac{s_\lambda(I_N)s_\lambda(\bpow^{(1)})s_\lambda(\bpow^{(2)})}{\left( s_\lambda(\bpow_\infty) \right)^2}
\ee
\[
 = \,\int \,\tau^{\rm B}_1\left( Z_1 Z_2 \right)  \,\prod_{i=1,2} \,
  e^{V(\tr Z^\dag_i,\,\bpow^{(i)})}\,d\Omega(Z_i,Z^\dag_i)
\]
If $\bpow^{(2)}=\bpow(\texttt{q},\texttt{t})$ with any given parameters $\texttt{q},\texttt{t}$,
then (\ref{3-points-integral}) is the hypergeometric BKP tau function. The same series over partitions may obtained
from integrals where $Z,Z^\dag$ are replaced either by normal matrices $M,M^\dag$ or by a pair of Hermitian matrices
(for these replacements see \cite{NO-2014}).

{\bf Example 3. Hermitian two-matrix model}.
The following
'projective analogue' of the well-known two matrix model is the BKP tau function
\[
\int \tau^{\rm B}_1(c M_2)  e^{\tr V(M_1,\bpow)+\tr (M_1 M_2)}dM_1dM_2 =
\sum_{\lambda}\,c^{|\lambda|} (N)_\lambda  s_\lambda(\bpow)
\]
where $M_1,M_2$ are Hermitian matrices.
Using results of \cite{GJ} we can show that it is a projective analogue of the generating function of the
so-called strictly monotonic Hurwitz numbers introduced by Goulden and Jackson. In the
projective case these numbers counts paths on Cayley graph of the symmetric group whose initial point is a given partition
while the end point is not fixed: we take the weighted sum over all possible end points, say, $\Delta$,  the weight is
given by $\chi(\Delta)$ of Lemma \ref{Hurwitz-down-Lemma}.

{\bf Example 4. Unitary matrices.} Generating series for projective Hurwitz numbers with arbitrary profiles
in $n$ branch points and restricted profiles in other points:

\be\label{multimatrix-unitary-RP2'}
\int e^{\tr (c U_1^\dag \dots U_{n+m}^\dag)}
\left(\prod_{i=n+1}^{n+m} \tau^{\rm B}_1(U_i) d_*U_i \right)
\left(\prod_{i=1}^{n} \tau^{\rm KP}_1(U_i,\bpow^{(i)}) d_*U_i \right)
=
\ee
\[
\sum_{d\ge 0}c^d \left( d! \right)^{1-m} \sum_{\lambda,\, |\lambda|=
d\atop \ell(\lambda)\le N}\, \left(\frac{{\rm dim}\lambda}{d!}  \right)^{2-m}
\left(\frac{s_\lambda(I_N)}{{\rm dim} \lambda} \right)^{1-m-n}
\prod_{i=1}^n \frac{s_\lambda(\bpow^{(i)})}{{\rm dim} \lambda}
\]

Here $\bpow^{(i)}$ are parameters. This series generate certain linear combination of Hurwitz numbers for base surfaces
with Euler characteristic $2-m,\,m\ge 0$.
The integral (\ref{multimatrix-unitary-RP2'}) is a BKP tau function in case the parameters
are specialized as $\bpow^{(i)}=\bpow(\texttt{q}_i,\texttt{t}_i),\,i=2,\dots,n$ with any values of
$\texttt{q}_i,\texttt{t}_i$,
and if in addition $m=1$. In case $n=1$ this BKP tau function may be viewed as an analogue of the generating function of
the so-called non-connected Bousquet-Melou-Schaeffer numbers
(see Example 2.16 in \cite{KazarianLando}).
In case $n=m=1$ we obtain the following BKP tau function
\[
\int \tau^{\rm B}_1(U_2)  e^{\tr V(U_1,\bpow)+\tr (cU_1^\dag U_2^\dag)}d_*U_1d_*U_2 =
\sum_{\lambda\atop \ell(\lambda)\le N}\,c^{|\lambda|}\frac{s_\lambda(\bpow)}{(N)_\lambda}
\]
If we compare this series with ones used in \cite{Goulden-Paquet-Novak},\cite{Harnad-2014} we can see that it is a projective analogue of the generating function of the
so-called weakly monotonic Hurwitz numbers. In the projective case it counts paths on Cayley graph whose initial point is a given partition
while instead of a fixed end point we consider the sum over the all possible end points $\Delta$, 
with a weight given by $\chi(\Delta)$ of Lemma \ref{Hurwitz-down-Lemma}.

{\bf Example 5. Integrals over complex matrices}.
A pair of examples. 
An analogue of Belyi curves generating function \cite{Zog},\cite{Chekhov-2014} is as follows
 (compare also to (\ref{S})):
\be
\sum_{ \Delta^{(1)},\dots,\Delta^{(k+1)}\atop \ell(\Delta^{k+1})=l}\,\frac{N^l}{d!}\,
H^{1,k+1}(d;\Delta^{(1)},\dots,\Delta^{(k+1)})\,
\prod_{i=1}^{k+1} \bpow^{(i)}_{\Delta^{(i)}}\,
=\sum_{\lambda}\,(N)_\lambda s_\lambda(\bpow^{(k+1)})\,
\prod_{i=1}^{k}\frac{s_\lambda(\bpow^{(i)})}{s_\lambda(\bpow_\infty)}
\ee
\be
=\,\int \tau_1^{\rm KP}(Z^\dag Z_1^\dag \cdots Z_k^\dag,\bpow^{(k+1)})\tau_1^{\rm B}(Z)d\Omega(Z,Z^\dag)
\,\prod_{i=1}^{k}\,\tau_1^{\rm KP}(Z_i,\bpow^{(i)})  d\Omega(Z_i,Z_i^\dag)
\ee

The series in the following example generates the projective Hurwitz numbers themselves where to get rid
of the factor $(N)_\lambda$ in the sum over partitions we use mixed integration over $\mathbb{U}(N)$ and over
complex matrices:
\be
\sum_{ \Delta^{(1)},\dots,\Delta^{(k+1)}}\,\frac{c^d}{d!}\,
H^{1,k+1}(d;\Delta^{(1)},\dots,\Delta^{(k+1)})\,
\prod_{i=1}^{k+1} \bpow^{(i)}_{\Delta^{(i)}}\,
=\sum_{\lambda,\,\ell(\lambda)\le N}\,c^{|\lambda|} s_\lambda(\bpow^{(k+1)})\,
\prod_{i=1}^k\frac{s_\lambda(\bpow^{(i)})}{s_\lambda(\bpow_\infty)}
\ee
\be
=\,\int \tau_1^{\rm KP}(c U^\dag Z_1^\dag \cdots Z_k^\dag,\bpow^{(k+1)})\tau_1^{\rm B}(U)d_*U \prod_{i=1}^k
\tau_1^{\rm KP}(Z_i,\bpow^{(i)}) d\Omega(Z_i,Z_i^\dag)
\ee
Here $Z,Z_i,\,i=1,\dots,k$ are complex $N\times N$ matrices and $U\in\mathbb{U}(N)$. As in the previous examples
one can specify all sets $\bpow^{(i)}=\bpow(\texttt{q}_i,\texttt{t}_i),\,i=1,\dots,k+1$ except a single one which in 
this case has the meaning of the BKP higher times.

\section*{Acknowledgements}

A.O. was partially supported by RFBR grant 14-01-00860 and by V.E.Zakharov's scientific school
(Program for Support of Leading Scientific Schools, grant NS-3753.2014.2) and also by
by the  Russian Academic Excellence Project '5-100'.
The work of S.N. was partially supported by RFBR grants 15-52-50041.
We thank first of all J. Harnad, A. Mironov and J. van de Leur for important remarks, and also
A. Zabrodin,  S. Loktev and I. Marshall for useful discussions.
Our special grates to L. Chekhov for the organization of the workshop on Hurwitz numbers (Moscow, May 2014) which
inspired us to do this work.

\appendix

 \section{Hirota equations for the BKP tau function with two discrete time variables.\label{N-n-BKP-Hirota}}

 The BKP hierarchy we are interested in was introduced in \cite{KvdLbispec}.
 In this paper the BKP tau function $\tau^{\rm B}(N,\bpow)$ does not contain the discrete variable $n$.
 We need in a slightly general version of BKP hierarchy which includes $n$ as the higher time parameter,
 see \cite{OST-I} and \cite{LeurO-2014}.
 Hirota equations for the tau functions $\tau^{\rm B}(N,n,\bpow)$ of this modified BKP hierarchy read
\bea\label{Hirota-N-n-BKP-OST'}
  \oint\frac{dz}{2\pi i}z^{N'-N-1}e^{V(\bpow'-\bpow,z)}
  \tau(N'-1,n,\bpow'-[z^{-1}])
  \tau(N+1,n+1,\bpow+[z^{-1}]) \nonumber\\
+ \oint\frac{dz}{2\pi i}z^{N-N'-3}e^{V(\bpow-\bpow',z)}
  \tau(N'+1,n+2,\bpow'+[z^{-1}])
  \tau(N-1,n-1,\bpow-[z^{-1}]) \nonumber\\
=
  \tau(N'+1,n+1,\bpow')
  \tau(N-1,n,\bpow)
- \frac{1}{2}(1-(-1)^{N'+N})
  \tau(N',n+1,\bpow'|g)\tau(N,n,\bpow)
\eea
 and
\bea\label{Hirota-N-BKP(deLeur)'}
  \oint\frac{dz}{2\pi i}z^{N'-N-2}e^{V(\bpow'-\bpow,z)}
  \tau(N'-1,n-1,\bpow'-[z^{-1}])\tau(N+1,n+1,\bpow +[z^{-1}]) \nonumber\\
+ \oint\frac{dz}{2\pi i}z^{N-N'-2}e^{V(\bpow-\bpow',z)}
  \tau(N'+1,n+1,\bpow'+[z^{-1}])\tau(N-1,n-1,\bpow'-[z^{-1}]) \nonumber\\
= \frac{1}{2}(1-(-1)^{N'+N})\tau(N',n,\bpow')\tau(N,n,\bpow)
\eea

 Here $\bpow=(p_1,p_2,\dots)$, $\bpow'=(p'_1,p'_2,\dots)$.
The notation $\bpow +[z^{-1}]$ denotes the set $\left( p_1+z^{-1},p_2+z^{-2}, p_3+z^{-3},\dots \right)$ and
$V$ is defined by (\ref{V}).

Equations (\ref{Hirota-N-BKP(deLeur)'}) are the same as in \cite{KvdLbispec} while equations
(\ref{Hirota-N-n-BKP-OST'}) relate tau functions with different discrete time $n$ and were written down
in \cite{OST-I} and \cite{LeurO-2014}.

 Taking $N'=N+1$ and all $p_i=p_i',\,i\neq 2$ in (\ref{Hirota-N-BKP(deLeur)'})
and picking up the terms linear in $p'_2-p_2$ we obtain (\ref{Hirota-elementary-1'}).
Taking $N'=N+1$ and all $p_i=p_i',\,i\neq 1$ in (\ref{Hirota-N-n-BKP-OST'})
and picking up the terms linear in $p'_1-p_1$ we obtain (\ref{Hirota-elementary-2'})

The relation of the BKP hierarchy to the two- and three-component KP hierarchy was established in \cite{LeurO-2014}.

\section{Hypergeometric BKP tau function. Fermionic formulae\label{fermionic-appendix}}
Details may be found in \cite{OS-2000, OST-I}.
Let
 $\{\psi_i$, $\psi_i^\dag$, $i \in \mathbb{Z}\}$ are Fermi creation and
annihilation operators that  satisfy the usual anticommutation relations and vacuum annihilation conditions
\[
[\psi_i, \ \psi_j]_+ = \delta_{i,j}, \quad \psi_i | n\rangle =
\psi_{-i-1} | n\rangle =0,\quad   i< n
 \]
In contrast to the DKP hierarchy introduced in \cite{JM} for the BKP hierarchy introduced in \cite{KvdLbispec}
one needs an additional Fermi mode  $\phi$ which anticommutes with each other
Fermi operator except itself: $\phi^2=\frac 12$, and
 $\phi|0\r=\frac{1}{\sqrt{2}}|0\r$, see \cite{KvdLbispec}. Then the hypergeometric BKP tau function introduced in
 \cite{OST-I} may be written as
 \[
 g(n)\tau^{\rm B}_r(N,n,\bpow) =
 \l n| e^{\sum_{m>0} \frac 1m J_m p_m}
 e^{\sum_{i < 0} U_i \psi_i^\dag \psi_i -\sum_{i \ge 0} U_i \psi_i\psi_i^\dag }
 e^{\sum_{i>j} \psi_i\psi_j\,-\sqrt{2}\,\phi \sum_{i\in\mathbb{Z}} \psi_i}|n-N\r =
 \]
 \be\label{hyper-via-U,r}
= \sum_{\lambda\atop \ell(\lambda)\le N} \,e^{-U_\lambda(n)} s_\lambda(\bpow)
 = g(n)
\sum_{\lambda\atop \ell(\lambda)\le N}  r_\lambda(n)s_\lambda(\bpow)
\ee
where  $J_m=\sum_{i\in\mathbb{Z}}\,\psi_i\psi^\dag_{i+m},\ m>0$,
$U_\lambda(n)=\sum_{i} U_{h_i+n}$,
$r(i)=e^{U_{i-1}-U_{i}}$
and
\bea
 e^{-U_0+\cdots -U_{n-1}}\quad {\rm if}\,\, n>0 \qquad\qquad\qquad \\
g(n)\,:=\,\l n|  e^{\sum_{i < 0} U_i \psi_i^\dag \psi_i -\sum_{i \ge 0} U_i \psi_i\psi_i^\dag }
|n\r\,= \qquad\qquad\qquad\qquad\qquad
1\quad {\rm if} \,\,n=0 \qquad\qquad\qquad
\label{g(n)}
\\
e^{U_{-1}+\cdots U_{n}}\quad {\rm if}\,\, n<0 \qquad\qquad\qquad
\eea
In (\ref{hyper-via-U,r}) the summation runs over all partitions whose length do not exceed $N$.
\br\label{DKPvsBKP}
Let us note that without the additional Fermi mode $\phi$ the summation range in (\ref{hyper-via-U,r}) does
include partitions with odd partition lengths. One can avoid this restriction by introducing a pair of DKP tau
functions which seems unnatural.
\er
Apart of (\ref{hyper-via-U,r}) the same series without the restriction $\ell(\lambda)\le N$ is the example
of the BKP tau function however it is related to the single value $n=0$, the $n$-dependence destroys the
simple form of such tau function, see \cite{OST-I}.

\end{document}